\documentclass[aps, prb, superscriptaddress, preprint, showpacs, amssymb, onecolumn]{revtex4}
\usepackage{amsmath}
\usepackage{amssymb}
\usepackage{amsthm}
\usepackage{graphicx}
\usepackage{enumerate}
\usepackage{color}

\begin{document}

\title{Design of a 3D photonic band gap cavity in a diamond-like inverse woodpile photonic crystal}

\author{L\'eon A. Woldering}\email{l.a.woldering@utwente.nl}
\affiliation{Transducers Science and Technology (TST), MESA+ Institute for
Nanotechnology, University of Twente, P.O. Box 217, 7500 AE
Enschede, The Netherlands}

\author{Allard P. Mosk}
\affiliation{Complex Photonic Systems (COPS), MESA+ Institute for
Nanotechnology, University of Twente, P.O. Box 217, 7500 AE
Enschede, The Netherlands}

\author{Willem L. Vos}\email{w.l.vos@utwente.nl; www.photonicbandgaps.com}
\affiliation{Complex Photonic Systems (COPS),
MESA+ Institute for Nanotechnology,
University of Twente,
P.O. Box 217,
7500 AE Enschede,
The Netherlands}

\pacs{42.50.Pq, 42.70.Qs}

\date{Prepared May 12th, 2014. Posted on arXiv.}

\begin{abstract}
We theoretically investigate the design of cavities in a three-dimensional (3D) inverse woodpile photonic crystal.
This class of cubic diamond-like crystals has a very broad photonic band gap and consists of two perpendicular arrays of pores with a rectangular structure.
The point defect that acts as a cavity is centred on the intersection of two intersecting perpendicular pores with a radius that differs from the ones in the bulk of the crystal.
We have performed supercell bandstructure calculations with up to $5 \times 5 \times 5$ unit cells.
We find that up to five isolated and dispersionless bands appear within the 3D photonic band gap.
For each isolated band, the electric-field energy is localized in a volume centred on the point defect, hence the point defect acts as a 3D photonic band gap cavity.
The mode volume of the cavities resonances is as small as 0.8 $\lambda^{3}$ (resonance wavelength cubed), indicating a strong confinement of the light.
By varying the radius of the defect pores we found that only donor-like resonances appear for smaller defect radius, whereas no acceptor-like resonances appear for greater defect radius.
From a 3D plot of the distribution of the electric-field energy density we conclude that peaks of energy found in sharp edges situated at the point defect, similar to how electrons collect at such features. This is different from what is observed for cavities in non-inverted woodpile structures.
Since inverse woodpile crystals can be fabricated from silicon by CMOS-compatible means, we project that single cavities and even cavity arrays can be realized, for wavelength ranges compatible with telecommunication windows in the near infrared.
\end{abstract}
\maketitle

\section{Introduction}
\hspace{\parindent}
Many efforts are currently proceeding in the blossoming field of nanophotonics to trap light in a tiny volume in space~\cite{Vahala2003, Novotny2006}.
Several classes of devices are pursued including micropillar and ring cavities~\cite{Armani2003, Gerard2003}, point defects in two-dimensional photonic crystals~\cite{Akahane2003, Hess2003pssa}, and plasmonic structures such as metallic antennas~\cite{Farahani2005, Taminiau2008}.
Nanophotonic resonators have many interesting potential applications, such as the trapping or slowing-down of photons~\cite{Vahala2003}, sensing~\cite{krio2002}, a controlled enhancement of spontaneous emission~\cite{Gerard1998}, as well as advanced cavity quantum electrodynamic control~\cite{Reithmaier2004, Yoshie2004N, Peter2005}.
Linear arrays of wavelength-scale optical cavities are pursued for their function as waveguides with tailored properties~\cite{yari1999, Yann2002}.

Of particular interest are cavities embedded in three-dimensional (3D) photonic crystals with a complete photonic band gap~\cite{Yablonovitch1987prl, John1987prl}.
In the frequency range of the band gap light is forbidden to exist throughout the crystal and for all polarizations, which notably leads to the inhibition of spontaneous emission~\cite{Leistikow2011PRL}.
By introducing a point defect into the crystal structure, the lattice symmetry is locally broken, and a resonance appears in the band gap~\cite{Yablonovitch1991PRL-2, Ozbay1995, vill1996, Joannopoulos2008}, in an analogy to localized electronic defect states in a semiconductor~\cite{Ashcroft1976, Economou2010}.
A 3D photonic band gap cavity is considered to be an ultimate tool to control light down to the single photon level for several reasons.
First, since the confinement is truly three-dimensional, there is no direction or dimension wherein the light will naturally leak as is the case in, \emph{e.g.}, a pillar or a 2D photonic crystal.
Second, since in photonic band gap crystals the imaginary part of the dielectric constant of the constituent materials is minimal, the absorption of light is minimal, allowing very long storage times of light.
Third, since a 3D photonic band gap effectively shields an embedded quantum system, such as an excited quantum dot, from vacuum fluctuations, an array of 3D cavities has great potential to control collective quantum systems including qubits~\cite{Vos2014Cambridge}.

It is a major challenge in nanotechnology to realize optical cavities in 3D crystals~\cite{brau2006, yan2007}, since a controlled deviation from the periodic crystal structure must be realized deep inside the nanostructure.
One demonstrated solution to this challenge are cavities in woodpile structures made with a layer-by-layer method~\cite{lin1999, Okano2002,ogaw2004, tand2011}.
In this method crystals are made by sequential stacking of layers where the central layer is modified to contain a point defect.
Unfortunately layer-by-layer stacking suffers from random fluctuations in the alignment.
As a result the width of the photonic band gap is limited, hence the density of the optical states in the gap becomes filled with undesired states, thereby limiting the cavity quality factor.
In a second method an optical cavity was proposed by an intriguing combination of a planar unit cell modulation (planar defect) and a waveguide (line defect)~\cite{tang2007_2, tang2011}.
Relevant and interesting methods to fabricate cavities in opal -and inverse opal- based photonic crystals has been reported in references~\cite{Ferr2004, Lee2002, rinn2007, rama2008}.

\begin{figure}
\centering
\includegraphics[width=0.5\columnwidth]{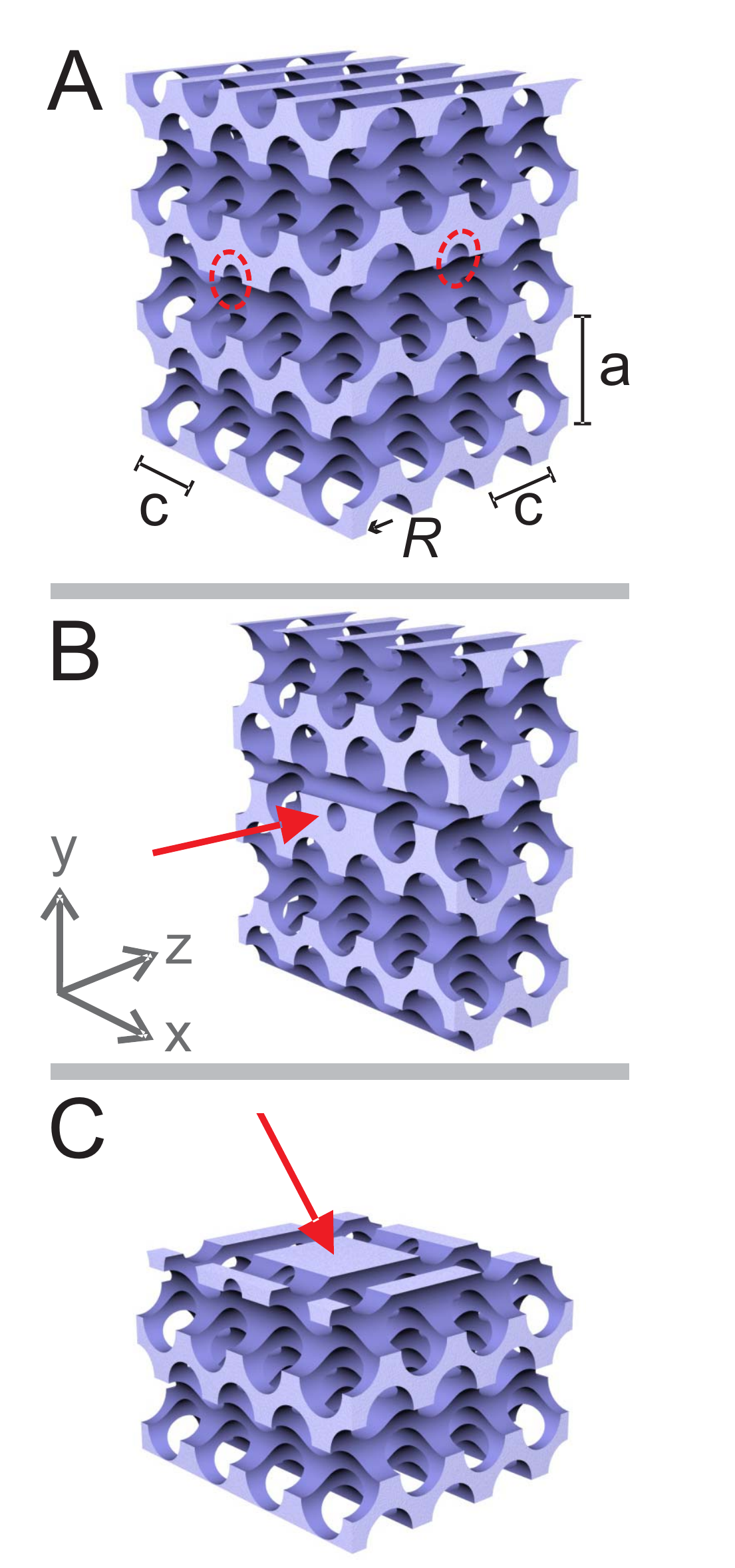}
\caption{(Color online) Structure of an inverse woodpile photonic crystal with a point defect.
(A) Oblique view of a section of the crystal with defect. The radius $R$ of the pores and the orthorhombic lattice parameters $a$ and $c$ are indicated.
The red circles emphasize two pores with a radius $R^`$ different from the bulk radius $R$.
Here the defect pores are smaller than the bulk pores: their radius $R^`$ is equal to 0.5~$R$.
(B) Vertical cross-section through the centre of a point defect cavity. The used x,y,z-coordinate system is shown in grey.
(C) Horizontal cross-section through the center of a point defect.
The two intersecting defect pores result in a region with an excess of high-index material, indicated by the arrows.}
\label{Schematicsofdefect}
\end{figure}

In this paper, we propose and investigate a straightforward approach to realize an optical cavity in an inverse woodpile photonic band gap crystal~\cite{Ho1994SSC}.
These photonic band gap crystals have a symmetry similar to how carbon atoms are arranged in a diamond crystal~\cite{Ho1994SSC}.
Diamond-like photonic crystals stand out for their broad band gaps~\cite{Maldovan2004}, as a result of which an embedded cavity is optimally shielded.
In addition, a broad photonic band gap offers robustness to unavoidable disorder and to inadvertent fabrication deviations~\cite{Li2000PRB, Woldering2009}.
Among the diamond-like crystals, the inverse woodpile stand out because they are relatively straightforward to fabricate by etching two perpendicular arrays of pores in a high-refractive index material such as silicon~\cite{Hillebrand2003JAP, Schilling2005APL, Tjerkstra2011JVST, Broek2012AFM}, as illustrated in Figure~\ref{Schematicsofdefect}.
Recent work on silicon inverse woodpile crystals has demonstrated the experimental signature of a broad 3D photonic band gap in reflectivity~\cite{Huisman2011PRB}, and a strong inhibition of spontaneous emission of embedded quantum emitters~\cite{Leistikow2011PRL}

\begin{figure}
\centering
\includegraphics[width=1.0\columnwidth]{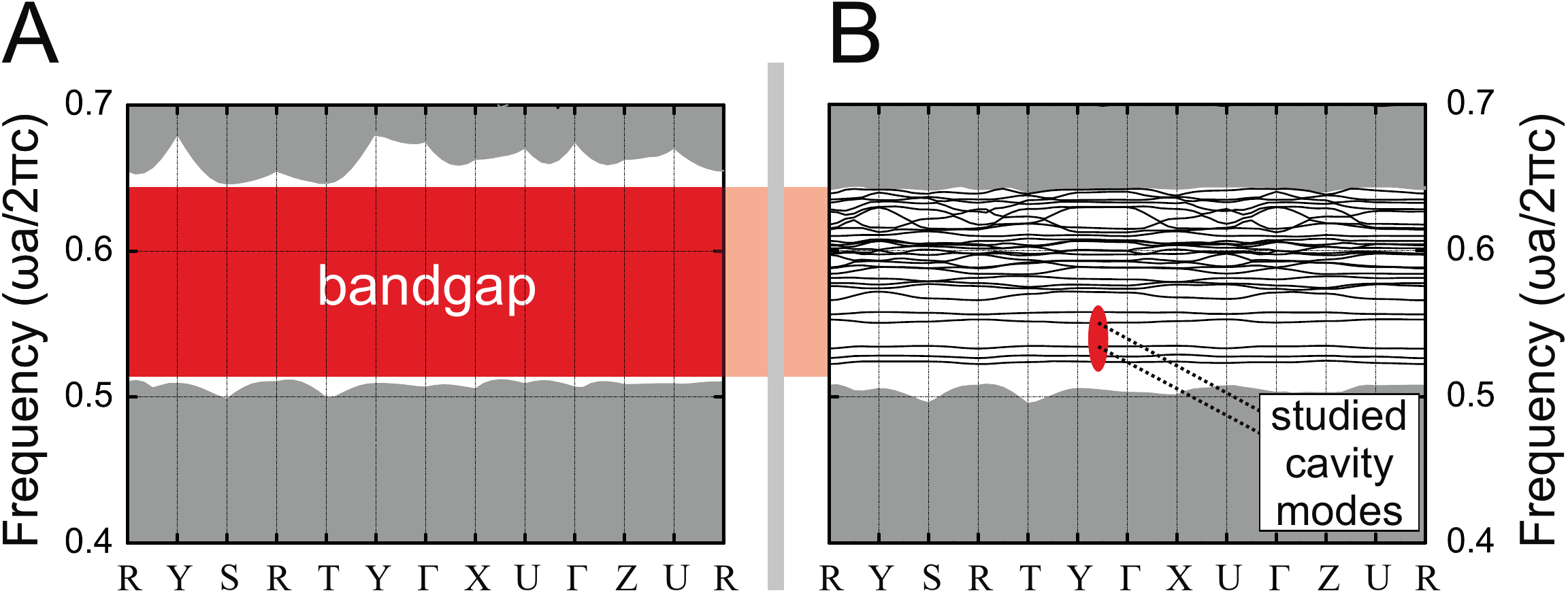}
\caption{(Color online)
(A) Bandstructure calculated for a perfect inverse woodpile crystal.
The dielectric constant is $\varepsilon_{\mathrm{Si}}$~=~12.1, typical for silicon.
The 3D photonic band gap is shown as a red bar between reduced frequencies 0.510 and 0.646.
For ease of illustration, all allowed frequencies outside the original band gap are depicted in grey.
(B) Bandstructure calculated for two intersecting defect pores with radius $R^`$~=~$0.5R$, using a 3$\times$3$\times$3 supercell.
The abscissa indicates the wave vector between the eight high-symmetry points of the Brillouin zone of the orthorhombic lattice~\cite{Woldering2009}.
Many bands appear within the original band gap.
The five indicated flat bands are isolated and appear to confine light as cavity resonances. }
\label{BS_3x3x3}
\end{figure}

\section{Structure of crystal and point defect}
\hspace{\parindent}Figure~\ref{Schematicsofdefect} illustrates the structure of an inverse woodpile photonic crystal.
The orthorhombic lattice constants are $a$ and $c$, and the radius of an unperturbed pore is $R$.
If the ratio of the lattice constants equals ${a}/{c}$~=~$\sqrt{2}$, the crystal is cubic with a diamond-like symmetry~\cite{Ho1994SSC}.
When the pore radius is tuned to ${R}/{a}$~=~$0.24$, the 3D photonic band gap has a very broad bandwidth as shown in Fig.~\ref{BS_3x3x3}(A), with a relative bandwidth $\Delta \omega / \omega_c = 25.3 \%$, with $\Delta \omega$ the frequency width of the band gap, and $\omega_c$ its center frequency~\cite{Hillebrand2003JAP, Woldering2009}.

We define three-dimensional cavities in these inverse woodpile crystals by introducing a point defect in the bulk of the crystal consisting of two intersecting perpendicular defect pores with a radius $R'$ that differs from the bulk pore radius $R$. A visualization of the defect is shown in Fig.~\ref{Schematicsofdefect}.
This intersection is the position where we expect the electric-field energy to be localized.
We will discuss the confinement of light in such a cavity and explore for which defect radius $R'$ optimal confinement is achieved, quantified by a minimal mode volume $V_\mathrm{mode}$ for the cavity resonances.
The benefits of the cavity proposed here are twofold: first, the required nanostructures can be realized with existing CMOS-compatible silicon nanofabrication techniques~\cite{Tjerkstra2011JVST, Broek2012AFM}, and second, no post-production steps are required to obtain a single cavity or even an array of cavities.

\section{Calculation method}
\hspace{\parindent}We have used the well-known MIT photonic bands package to calculate the photonic bandstructures (frequency $\omega$ versus wave vector $\vec{k}$) and the spatial electric-field energy density $\varepsilon |E|^2 (\vec{r})$ distributions using the plane-wave expansion~\cite{Johnson2001aa}.
To define an inverse woodpile crystal, an orthorhombic unit cell is used, as shown in Fig.~\ref{Schematicsofdefect}.
Throughout this paper, the dielectric constant is taken to be $\varepsilon_{\mathrm{Si}}$~=~12.1, typical for silicon.
More details on plane-wave calculations on inverse woodpiles are given in Ref.~\cite{Woldering2009}, and in appendix~\ref{appendix} we discuss the resolution of the present calculations.

To introduce a point defect as a cavity and increase its surrounding unperturbed volume, we define the crystal by means of a supercell~\cite{Meade1993}.
Since a plane-wave expansion assumes the structure under study to be infinitely extended, the supercell is replicated infinitely in all three dimensions.
Since the supercell under consideration has no surrounding vacuum, a limitation of the method is that a cavity quality factor cannot be calculated precisely.
To verify that the supercell method yields correct results, we have compared the results for a $3 \times 3\times 3$ supercell on a perfect crystal without point defect to the results obtained with a conventional single unit cell~\cite{Hillebrand2003JAP, Woldering2009}.
We found that these calculations agree well, see appendix~\ref{appendix}, thus validating the supercell method.

A $3 \times 3\times 3$ supercell consists of a total of $27$ orthorhombic unit cells.
The two perpendicular defect pores with a different radius $R'$ are defined across the entire supercell, that is, in two neighboring unit cells in each direction.
The resulting point defect is centered at the intersection of the two defect pores.
Therefore, in the infinite structure considered in plane-wave calculations, a defect occurs every three unit cells in each dimension.
The infinite repetition of cavities means \textit{de facto} that we are calculating the properties of an array of weakly coupled cavities.
It is reasonable to expect that with a larger supercell, the properties for a single cavity are approached.
Therefore, we have also performed time-intensive calculations with $5 \times 5 \times 5$ supercell, containing $125$ unit cells, where the point defect is repeated every five unit cells in each dimension.

\section{Dispersionless bands and localized field energy \label{section:field_distrib}}
\hspace{\parindent}
Using a $3 \times 3\times 3$ supercell, we have calculated the bandstructures of an inverse woodpile crystals with two intersecting defect pores.
In Figure~\ref{BS_3x3x3}(B) the bandstructure is shown for $R'$~=~$0.5R$.
A multitude of bands appears within the photonic band gap of the perfect crystal.
Of particular interest are the five lowest bands in the band gap.
Each of these bands is isolated in frequency from all other bands.
Since each band has nearly the same frequency $\omega_{i} (i = 1...5)$ at all wave vectors $\vec{k}$, it is reasonable by Fourier arguments that the frequency $\omega_{i}$ is localized in space $\vec{r}$.
Therefore, the five bands likely confine light, as cavity resonances.
We now turn to a detailed study of the five cavity resonances and their development as a function of defect radius $R'$.

\begin{figure}
\centering
\includegraphics[width=1.0\columnwidth]{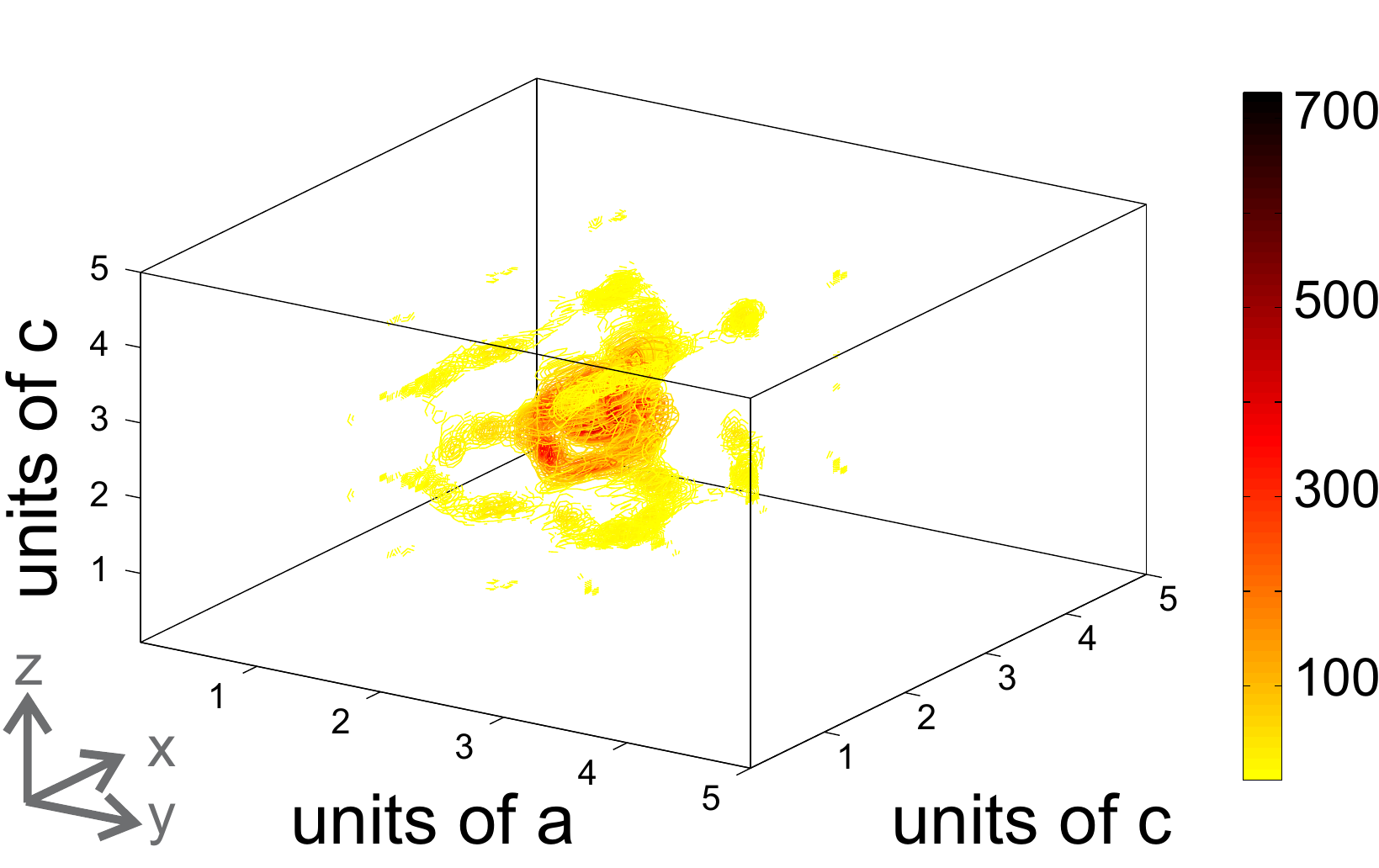}
\caption{(Color online)Three-dimensional electric-field energy density distribution of resonance 3 calculated on a 5$\times$5$\times$5 supercell with $R^`$~=~$0.5R$.
The cells are indicated along the axes in units of lattice spacings $a$ and $c$.
The used coordinate system is shown in grey.
For clarity, cells with very low energy densities ($\varepsilon |E|^2<2$) have been omitted.
}
\label{5x5x5_band-3_ED}
\end{figure}

Figure~\ref{5x5x5_band-3_ED} shows the electric-field energy distribution for the 3rd dispersionless band in the band gap at reduced frequency $(\omega a / 2\pi c) = 0.53$, with $c$ the speed of light.
For clarity, very low energy densities ($\varepsilon |E|^2 < 2$) have been omitted \footnote{Omitting energy densities less than ($\varepsilon |E|^2<2$) corresponds to a removal of only 7.5\% of the total electric-field energy density in the plots.
The remaining energy density of $92\%$ that is visualized occupies as little as $4\%$ of the total volume of the supercell.}.
A strong concentration of the electric-field energy density is observed at the center of the supercell, where the two defect pores intersect.
The confinement of the electric field energy further confirms that the five dispersionless bands in Figure~\ref{BS_3x3x3}(B) are cavity resonances.
Additionally, in Figure~\ref{5x5x5_band-3_ED}, there is no significant energy density along any of the two defect pores.
This result indicates that the defect pores do not act as waveguides through which light leaks out of the cavity.
At this point we conclude that a point defect in an inverse woodpile crystal centred on two intersecting defect pores yields a 3D photonic band gap cavity.

\begin{figure}
\centering
\includegraphics[width=0.75\columnwidth]{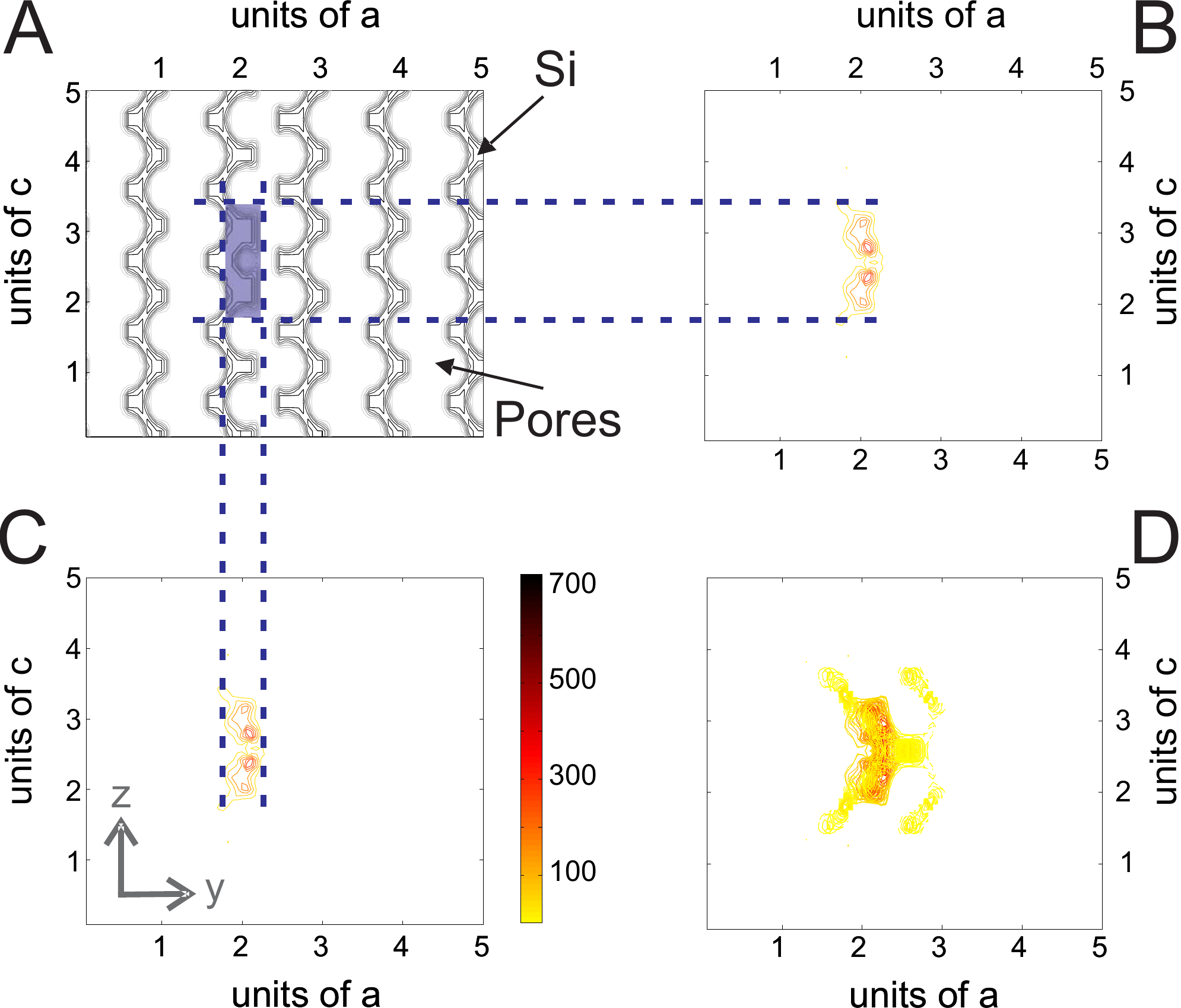}
\caption{(Color online)(A) Cross-section of the distribution of high-index material through the center of the defect, perpendicular to a 100 direction.
The axes indicate the extent of the supercell in units of lattice spacings $a$ and $c$.
The arrows point to an area with silicon and to a pore.
The two intersecting defect pores result in a region with an excess of high-index material, visible at the center.
(B) and (C) are identical cross sections of the electric-field energy density distribution $\varepsilon |E|^2$.
The coordinate system is shown in grey in panel (C).
The blue dashed lines indicate the overlap with dielectric distribution in (A).
The blue section in (A) indicates that the high energy density is found at the point defect.
(D )Projection of the electric-field energy density distribution through the whole supercell.
Multiple periodically spaced cross sections have been taken through the entire supercell and projected on top of each other.
All calculations were performed on a $5 \times 5 \times 5$ supercell with $R'~=~0.5R$.
The color bar in panel C also holds for panels B and D.
}
\label{100compare-eps-ED}
\end{figure}

\begin{figure}
\centering
\includegraphics[width=0.75\columnwidth]{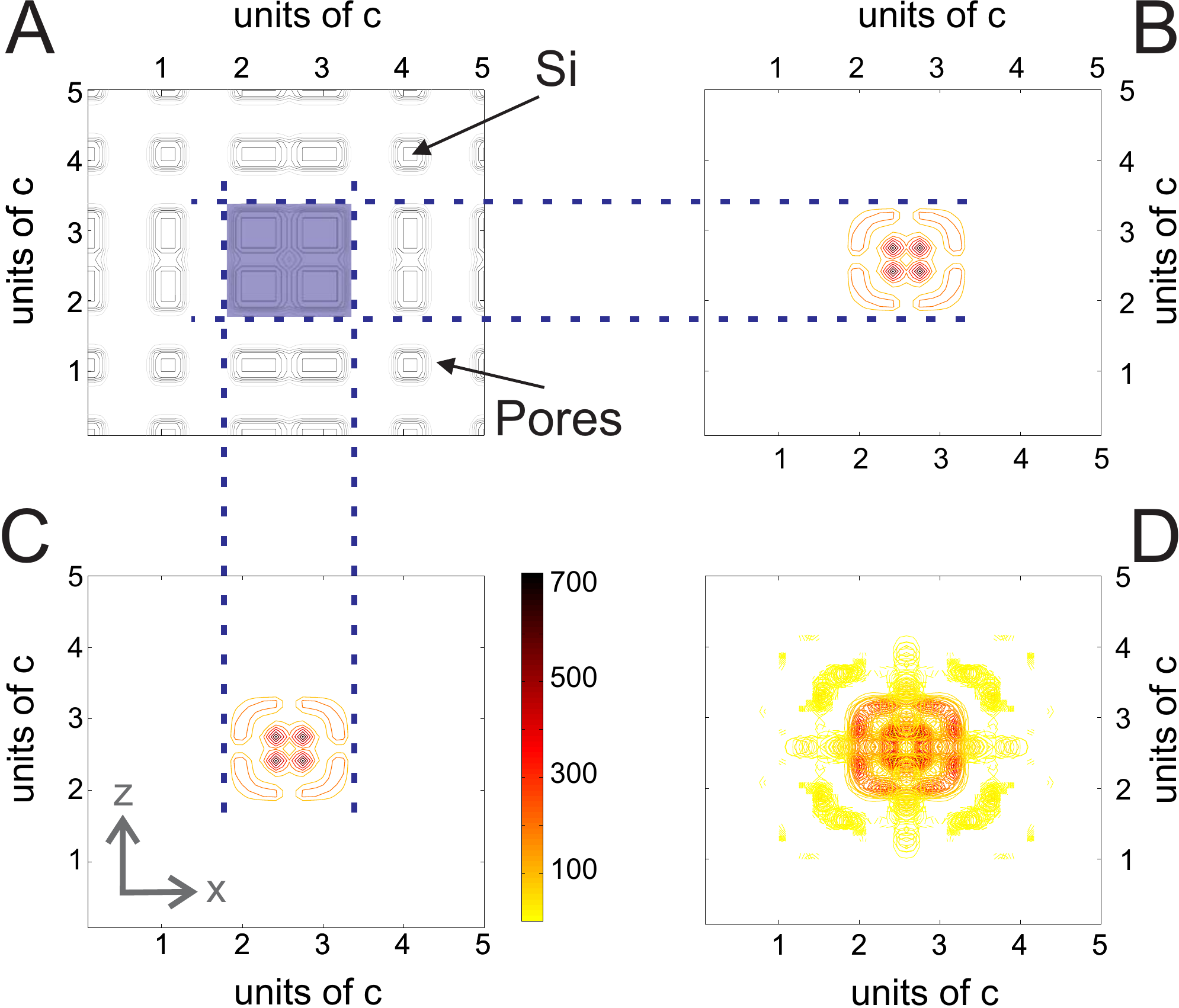}
\caption{(Color online)(A)Cross-section of the distribution of high-index material taken through the center of the defect, perpendicular to a 010 direction.
For clarity, arrows indicate an area with silicon and with a pore.
The two intersecting defect pores result in a region with an excess of high-index material, visible at the center of the panel.
(B) and (C) are identical cross sections of the electric-field energy density distribution $\varepsilon |E|^2$. By means of the blue dashed lines, the regions of high energy are traced back to the material distribution.
The blue section in (A) indicates that the high energy density is found at the point defect.
The coordinate system is shown in grey in panel (C).
(D)Projection of the electric-field energy density distribution through the whole supercell.
For clarity, very low energy densities ($\varepsilon |E|^2<2$) have been omitted.
Multiple periodically spaced cross sections have been taken through the entire supercell and projected on top of each other.
All calculations were performed on a $5 \times 5 \times 5$ supercell with $R'~=~0.5R$.
The cells are indicated along the axes in units of lattice spacings $a$ and $c$.
The views in this Figure are taken perpendicular to both pore directions.
The color bar in panel C is valid for panels B to D.
}
\label{010compare-eps-ED}
\end{figure}

In Figures 4 and 5 we show different cross-sections of the photonic crystal with defect, taken perpendicular to the 100 and 010 directions, respectively.
Figures~\ref{100compare-eps-ED}(A) and \ref{010compare-eps-ED}(A) show a cross-section of the distribution of high-index of refraction material in the supercell.
The pores can be easily identified, as well as the silicon backbone of the crystal structure.
Panels (B) and (C) are identical cross-sections of the calculated distribution of the electric-field energy density and used to compare the positions of high energy with the distribution of high-index material in panel (A).
The images confirm that at the position of the point defect, high electric-field energy densities are present, similar to what is described for woodpile-like structures~\cite{Joannopoulos2008}.
Taking a closer look at these Figures, we see that high densities of electric-field energy are present in a number of sharp tips in the point defect region. It appears that light is collected there, similar to the electric field enhancement seen in the lighting rod effect and used in field emission cathodes.

\section{Optimal pore radius \label{Optimaldefectradius}}

\begin{figure}
\centering
\includegraphics[width=0.75\columnwidth]{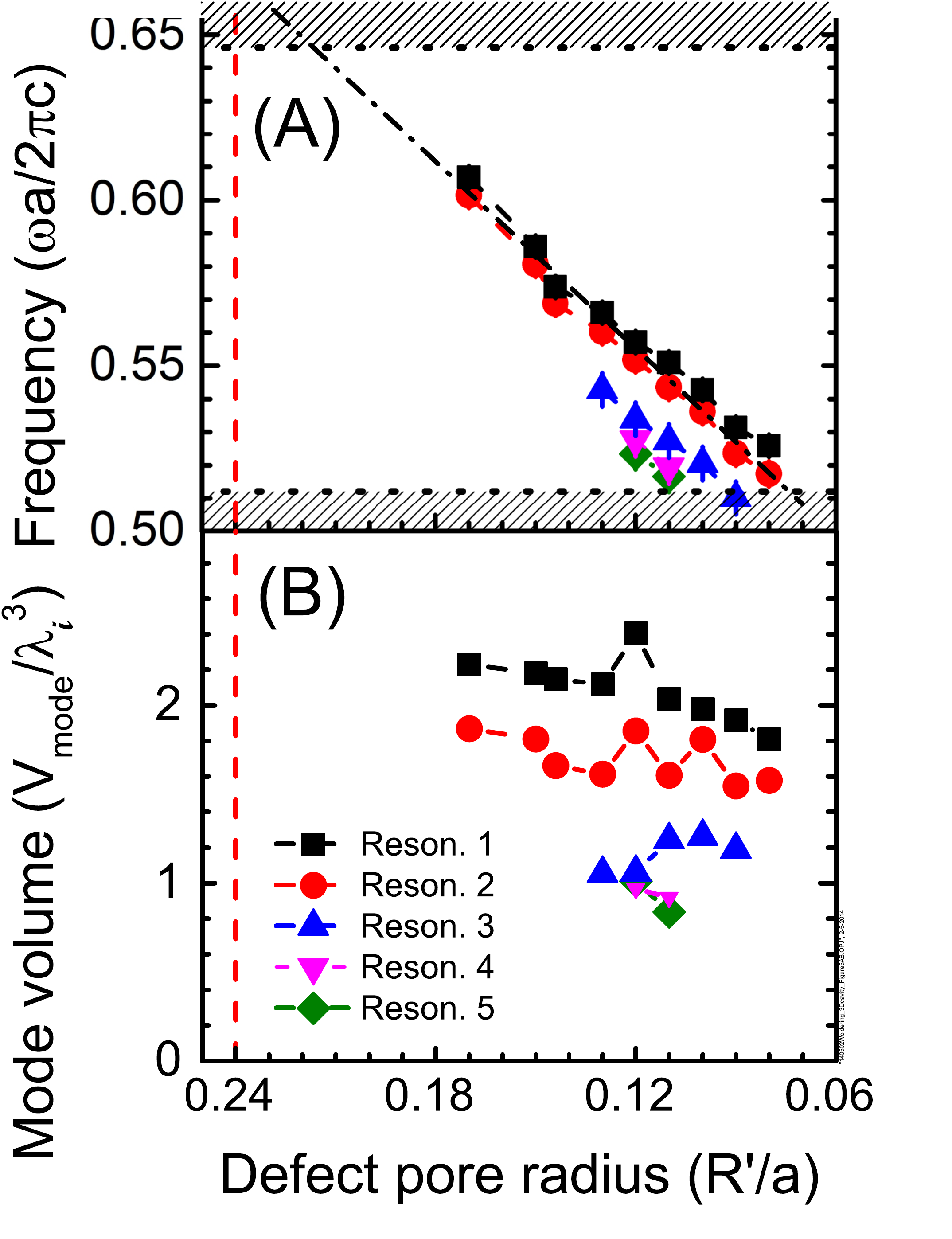}
\caption{\label{vandwbands}
(Color online)(A) Resonance frequencies in the photonic band gap versus defect pore radius $(R'/a)$.
The horizontal dashed line at reduced frequency 0.512 is the lower band edge and at reduced frequency 0.646 the upper band edge.
The red vertical dashed line at $(R'/a)=0.24$ is the perfect-crystal limit.
The black dashed-dotted line is a linear extrapolation of the first resonance frequency.
(B) Mode volume normalized to ${\lambda_\mathrm{mode}}^3$ as a function of defect pore radius.
The resonances 3, 4, and 5 have the smallest mode volumes.
Both panels show results for a $3 \times 3 \times 3$ supercell.
}
\end{figure}

\hspace{\parindent}To explore the conditions for confinement of light, we have varied the radius $R'$ of the two intersecting defect pores in the inverse woodpile crystals.
Figure~\ref{vandwbands}(A) shows the evolution of the frequency of each cavity resonance in the photonic band gap.
The frequency of a resonance is taken as the average between the lowest and highest frequency of each dispersionless band shown in Figure~\ref{BS_3x3x3}.
For convenience, we plot the ordinate with decreasing defect-pore radius, to obtain an ordinate that increases with the volume of the point defect, and $R'/a=0.24$ corresponds to a perfect crystal.
A defect radius of $R'/a = 0$ corresponds to zero defect pore radius, \emph{i.e.}, defect pores that are completely filled with high-index material.
Figure~\ref{vandwbands}(A) shows that the first two resonances appear near reduced frequency $(\omega a/2\pi c) = 0.6$ at a defect pore radius of $R'/a = 0.18$.
Here $c$ is the speed of light.
At a defect pore radius of $R'/a = 0.13$, a 3rd resonance appears below the first two, and even a 4th and a 5th resonance.
The resonances vanish in the ``valence bands" below the band gap at defect radius near $0.07$.
Thus, there are no donor resonances in the band gap anymore in the limit of completely filled defect pores $(R'/a = 0)$.

With increasing defect volume, the frequencies of all resonances decrease, consistent with results for increasing defects in 2D arrays of rods~\cite{vill1996, Joannopoulos2008} and in 3D direct woodpile structures~\cite{Okano2002}.
The decrease is physically intuitive since the resonances derive from an increasing volume of high refractive-index material.
By extrapolating the resonance frequencies to small defect volume, it is apparent that the resonances have split off from the top of the photonic band gap.
Based on the analogy between a photonic band gap and a semiconductor band gap, bands above the photonic band gap are referred to as ``conduction bands".
A bound state splitting off from a semiconductor's conduction band is a donor level, hence the observed resonances in the photonic band gap are referred as ``donor resonances"~\cite{Yablonovitch1991PRL-2, Joannopoulos2008}.

It is remarkable that the appearance of donor resonances in Figure~\ref{vandwbands}(A) occurs at a considerable threshold in defect radius.
Such a threshold is attribute to the fact that a certain minimum dielectric volume is required to sustain a standing wave in 3D.
As a result, the resonances appear at frequencies deep into the gap, as ``deep donors".
This behavior of donors in inverse woodpile crystals differs markedly from the occurrence of ``shallow" donor resonances in direct woodpile crystals~\cite{Okano2002}, and in fcc crystals with non-spherical atoms~\cite{Yablonovitch1991PRL-2}.
We surmise that the difference is a result of the field distribution of the cavity resonances.
In section~\ref{section:field_distrib} above, we have seen that in the inverse woodpile cavity the field maximum appears on the sharp corners of the dielectric.
Such sharp corners only appear when the defect radius differs considerably from the unperturbed radius, corresponding to a large detuning from the upper band edge, hence a ``deep donor".
In contrast, the cavity field distributions for the structures in Refs.~\cite{Yablonovitch1991PRL-2, Okano2002} are nearly completely localized in the additional high-index material.
Therefore, the cavity resonances appear for a smaller volume of additional material, and thus at smaller detuning from the upper band edge, corresponding to ``shallow donors".

In Figure~\ref{vandwbands}(B) the normalized mode volume ${V_\mathrm{mode}}/{\lambda_\mathrm{i}^3}$ is presented, determined as described in appendix~\ref{appendixVmode}.
For an optical cavity a small mode volume is desirable for strong confinement of light~\cite{Vahala2003, Novotny2006}.
The smallest mode volumes are observed for resonances 3, 4, and 5, with volumes of about ${V_\mathrm{mode}} = \lambda_\mathrm{i}^3$ at ${R'}/{a}~=~0.12$.
At ${R'}/{a}~=~0.11$, resonance 5 even has a mode volume as small as ${V_\mathrm{mode}} = 0.8 \lambda_\mathrm{i}^3$.
We note that resonance 3 appears to be of particular interest since it is isolated from the other resonances by the largest frequency gap.
If we combine this frequency isolation with the good confinement in real space as gauged by the small mode volume, we conclude that a defect radius of ${R'}/{a}~=~0.12$ is optimal, corresponding to an optimal defect pore radius $R'~=~0.5R$.
In the next section we investigate the properties of the resonances at this optimal condition in more detail by intensive computations on a large $5 \times 5 \times 5$ supercell.

\begin{figure}
\centering
\includegraphics[width=1.0\columnwidth]{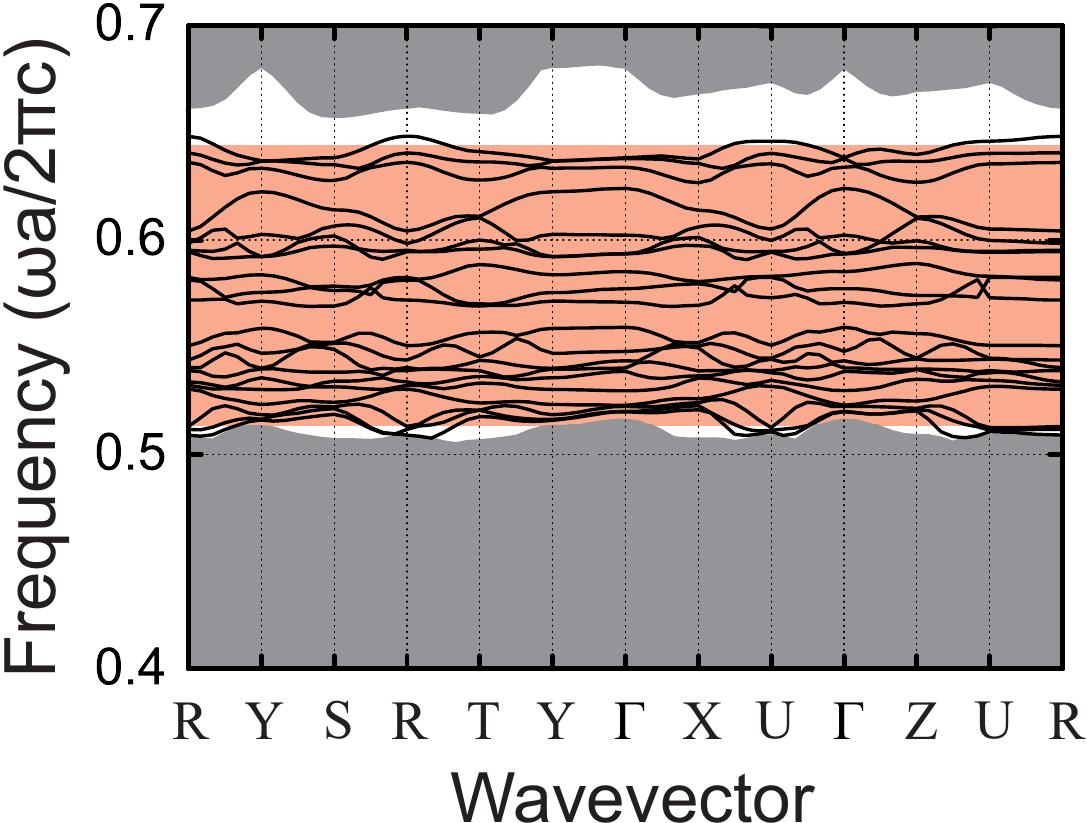}
\caption{(Color online) Photonic bandstructure calculated for an inverse woodpile crystal in which the two intersecting defect pores have a radius $\frac{R'}{a}$~=~0.35 larger than the bulk radius $R$.
No single isolated flat bands are observed.
There are isolated sets of 3 bands that may correspond to coupled resonances with low confinement.
The photonic band gap of a perfect crystal is shown as a red bar.}
\label{BS_035}
\end{figure}

To investigate whether inverse woodpile crystals also sustain ``acceptor resonances", we have performed calculations for defect pores with radii larger than the unperturbed pores ($R' > R$).
Figure~\ref{BS_035} shows a representative bandstructure calculated for ${R^`}/{a}~=~0.35$.
Intriguingly, no isolated resonances are observed.
Near the upper band gap edge, there is an isolated set of three bands near reduced frequency $(\omega a/2 \pi c) = 0.65$, with $c$ the speed of light.
A similar set of three bands is seen near the center of the band gap near reduced frequency $0.58$.
Since the individual bands within each triplet cross each other and do not form avoided crossings, it is likely that the field profiles of each band are orthogonal.
We consider the hypothesis unlikely that each band would form a separate localized resonance, since each band varies considerably in frequency.
We conclude that at large defect pore radii ($R' > R$) no acceptor resonances appear, in contrast to the well-confined donor resonances for small defect pore radii.

\begin{figure}
\centering
\includegraphics[width=1.0\columnwidth]{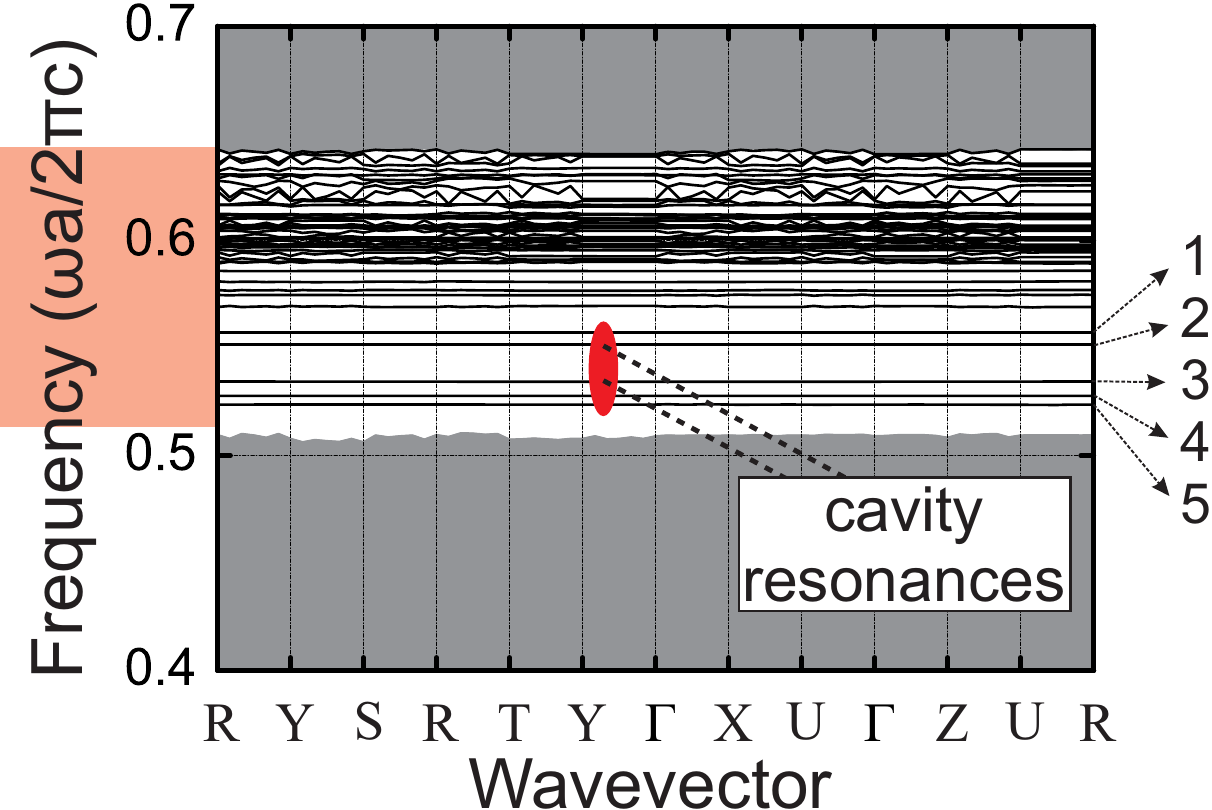}
\caption{(Color online) Calculated bandstructure of an inverse woodpile photonic crystal that is defined by a 5$\times$5$\times$5 supercell. In this structure two intersecting defect pores are present with a radius $R^`$~=~$0.5R$, similar to the results shown in Figure~\ref{BS_3x3x3}. The studied cavity resonances are labeled 1 to 5.}
\label{BS_5x5x5}
\end{figure}

\section{Resonance choice and practical realization \label{section-ED}}
\hspace{\parindent}
Using a 5$\times$5$\times$5 supercell, extensive calculations were performed on the optimal cavity condition identified in the previous section.
The bandstructure is shown in Figure~\ref{BS_5x5x5} for $R' = 0.5 R = 0.12 a$.
Again, five dispersionless bands appear near the bottom of the band gap.
The five bands are even flatter compared to the results for the $3 \times 3 \times 3$ supercell.
This can be rationalized with tight-binding arguments from solid-state physics, since bound states of atoms with an increasing density result in bands with an increased dispersion~\citep{Ashcroft1976, Economou2010}.
Therefore, the increased supercell calculation confirm that the cavity resonances are increasingly confined.

\begin{figure}
\centering
\includegraphics[width=0.75\columnwidth]{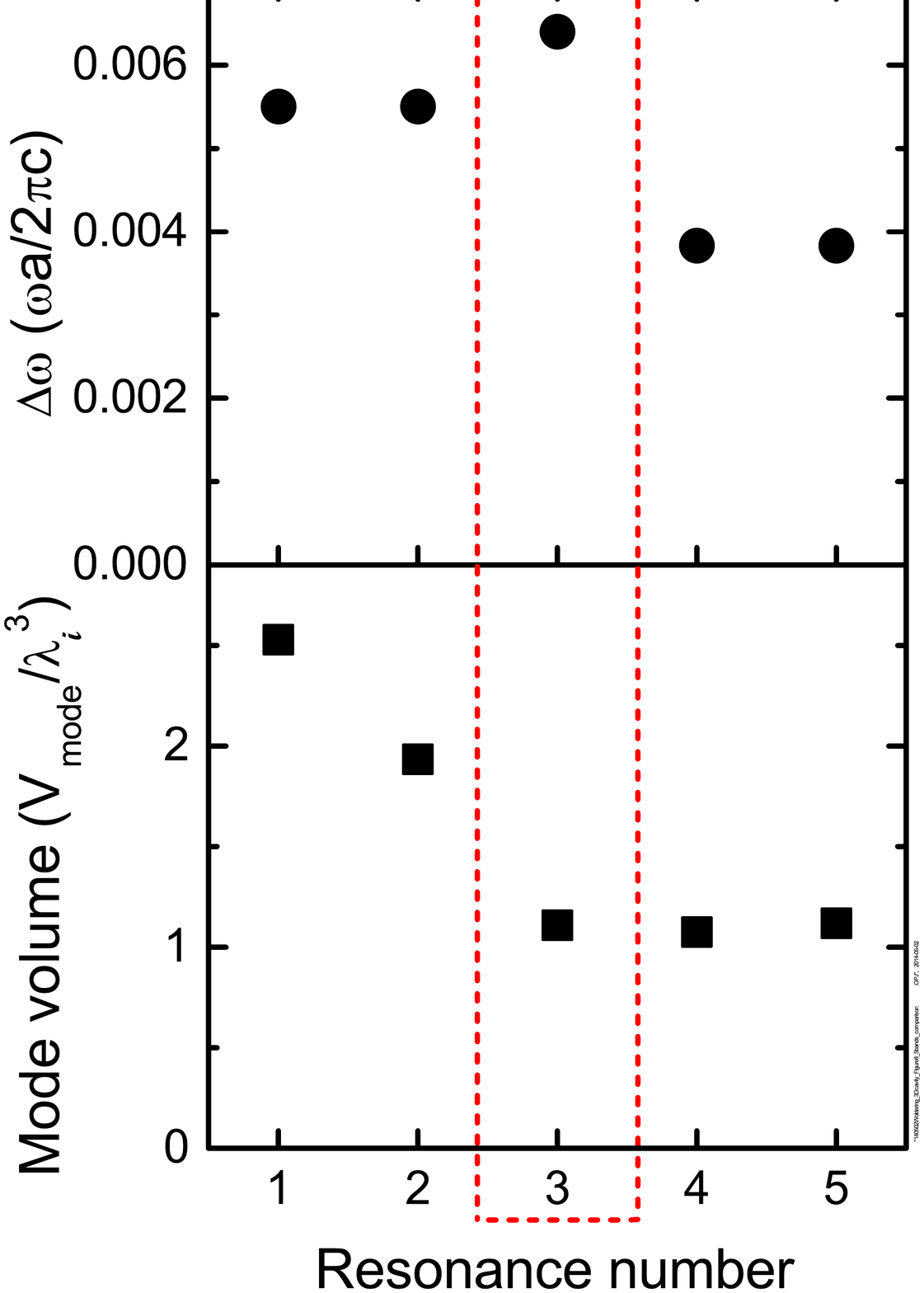}
\caption{(Color online) (A) Minimum separation to adjacent bands $\Delta\omega$ of the five resonances (5$\times$5$\times$5 supercell calculation for $\frac{R^`}{a}$~=~0.12).
We find the largest separation in frequency space for resonance 3.
(B) Calculated relative mode volume of the 5 resonances.
We observe the smallest mode volumes for resonances 1 to 3.
Based on these results resonance 3, indicated by the red dashed box, is most promising for strong confinement of light.}
\label{5bands_comparison}
\end{figure}

A detailed comparison of the five cavity resonances is presented in Figure~\ref{5bands_comparison}.
The potential of each resonance is gauged by two figures of merit:
\begin{enumerate}
\item{The isolation in frequency space, taken as the minimum separation to adjacent bands $\Delta\omega$.
A large separation to an adjacent band corresponds to a better confinement (or less leakage), and facilitates a selective optical addressing of a single resonance in experiments and applications.}
\item{The mode volume normalized to the wavelength cubed, where a smaller mode volume indicates a stronger spatial confinement of the light.}
\end{enumerate}

Figure~\ref{5bands_comparison}(A) confirms that the minimum separation to adjacent bands of resonance 3 is largest, with 1 and 2 having the smallest separation $\Delta\omega$.
Furthermore, Figure~\ref{5bands_comparison}(B) shows that the mode volumes of resonances 1 to 3 are the smallest.
The mode volumes are near $\frac{V_\mathrm{mode}}{\lambda_\mathrm{mode}^3}~=~1$, indicating that light is strongly confined.
The mode volumes of resonances 4 and 5 are significantly greater.
When combining these figures of merit for mode volume and frequency isolation, we conclude that resonance 3 has the best potential to confine light and to be selectively addressed.

We propose to pursue the fabrication of inverse woodpiles with embedded optical cavities by modifying our existing CMOS compatible manufacturing techniques.
In our realized silicon photonic crystals with typical structural properties (lattice spacings $a~=~680$ nm, $c~=~492$ nm,  and $\frac{R}{a}~=~0.24$) the frequency of the target resonance corresponds to a wavelength near 1270 nm.
This wavelength is in the telecommunication O-band, which makes these photonic band gap cavities relevant for applications.
By slightly tuning the lattice parameters $a$ and $c$, and the radii $R$ and$R'$, cavities can be made that operate in the C-band near 1550nm.
Furthermore, it is noted that the present cavity design is also relevant for inverse woodpile crystals made from alternative high-index materials, such as GaAs, GaP, or TiO$_{2}$.
These materials would even allow the realization of photonic band gap cavities at visible wavelengths.

\section{Conclusions}
\hspace{\parindent}We have performed supercell calculations on inverse woodpiles which contain two intersecting pores that have a different radius compared to the other pores in the crystal.
Our calculations show that isolated flat bands appear in the photonic band gap for defect radii smaller than the bulk radius, corresponding to donor levels in the band gap.
We have shown that the electric-field energy is concentrated about the center of the point defect, characteristic of a resonant optical cavity.
Despite the presence of the two defect pores, there are no preferential pathways for leaking of light of the cavity along each separate defect pore.

We have investigated five cavity resonances and found that the 3rd resonance at reduced frequency $(\omega a / 2 \pi c)$~=~0.534 is most promising for confinement.
We report a smallest mode volume of around 0.8 cubic wavelengths, typical of a strong spatial confinement of light.
By varying the radius of the defect pores, we have determined that a defect radius of $\frac{R^`}{a}$~=~0.12 gives the most optimal light confinement.
Finally we have discussed a practical method to realize the 3D photonic band gap cavities proposed here.

\section*{Acknowledgements}
\hspace{\parindent}The authors acknowledge Bas Benschop for assistance in setting up the computer on which the calculations were performed. This work was supported by the Stichting Fundamenteel Onderzoek der Materie (FOM) that is financially supported by the Nederlandse Organisatie voor Wetenschappelijk Onderzoek (NWO), by the Dutch Technology Foundation STW, which is part of the Netherlands Organisation for Scientific Research (NWO), and which is partly funded by the Ministry of Economic Affairs, by ERC grant 279248 to APM, and by a VENI fellowship (Technical Sciences) by the Netherlands Organisation for Scientific Research (NWO) to LAW.

\appendix

\begin{figure}
\centering
\includegraphics[width=1.0\columnwidth]{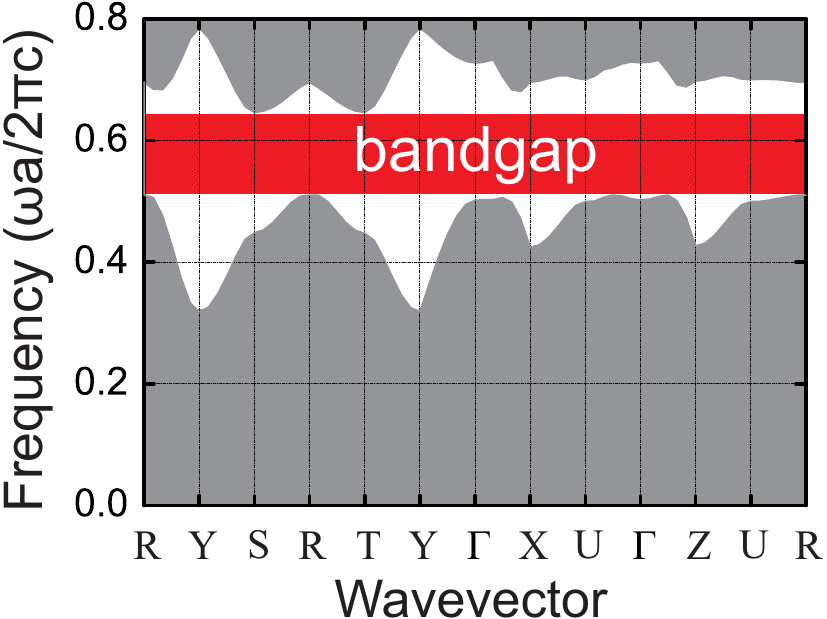}
\caption{(Color online) Calculated bandstructure of a perfect inverse woodpile photonic crystal.
Wave vector is shown between the eight high-symmetry points of the Brillouin zone of the orthorhombic lattice.
The grid resolution of this calculation is $12 \times 17 \times 12$.
Bands 1 to 8 are calculated for 73 k-points.
The dielectric constant is $\varepsilon_{\mathrm{Si}}$~=~12.1, typical for silicon.
A broad band gap with a relative bandwidth of $(\frac{{\Delta\omega}}{{\omega}}) $~=~23.5\% is seen between reduced frequencies 0.510 and 0.646, and indicated by the red bar.
For convenience, all allowed bands are depicted in grey.}
\label{BS_no_defect_1x1x1}
\end{figure}

\section{Grid resolution and supercells \label{appendix}}
\hspace{\parindent}
The MIT photonic bands program defines the unit cell with a certain resolution, the so-called grid resolution.
In general it is preferred to use a grid resolution that is as high as possible.
In our earlier work~\cite{Woldering2009}, we used a grid resolution of $68 \times 96 \times68$.
For the calculations in this paper, however, we have found that this resolution is too large in view of computer memory and time constraints.
This is due to the fact that in a supercell calculation this resolution is taken for each constituent cell.
Therefore we had to reduce the grid resolution to keep calculation times tractable.
Even then, the resulting computation time for the 5$\times$5$\times$5 supercell calculation was multiple months.

In Figure~\ref{BS_no_defect_1x1x1} a bandstructure is shown for a perfect inverse woodpile photonic crystal calculated with a resolution limited to $12 \times 17 \times 12$.
The band gap is indicated by the red bar.
The band edges and the width of the band gap compare well to those found in our earlier calculations~\cite{Woldering2009} with a higher grid resolution of $68 \times 96 \times68$, see Table~\ref{Table1}.
Consequently we conclude that the results with a lower grid resolution are sufficiently accurate for the present study.

\begin{table} \centering
\caption{Comparison of calculated band gap edges and band gap widths of the calculations with two different grid resolutions and supercell definitions.
\label{Table1}}
\begin{tabular}[t]{|c|c|c|c|}
\hline \textit{Grid resolution} & \textit{Supercell size} & \textit{Lower- and upper band gap edges} [$\omega$a/2$\pi$c] & \textit{Relative band gap width} [\%] \\
\hline 68$\times$96$\times$68 & 1$\times$1$\times$1 & 0.492~~~~~0.635 & 25.2 \\
\hline 12$\times$17$\times$12 & 1$\times$1$\times$1 & 0.510~~~~~0.646 & 23.5 \\
\hline 12$\times$17$\times$12 & 3$\times$3$\times$3 & 0.512~~~~~0.646 & 23.3 \\
\hline
\end{tabular}
\end{table}

\begin{figure}
\centering
\includegraphics[width=4.0in]{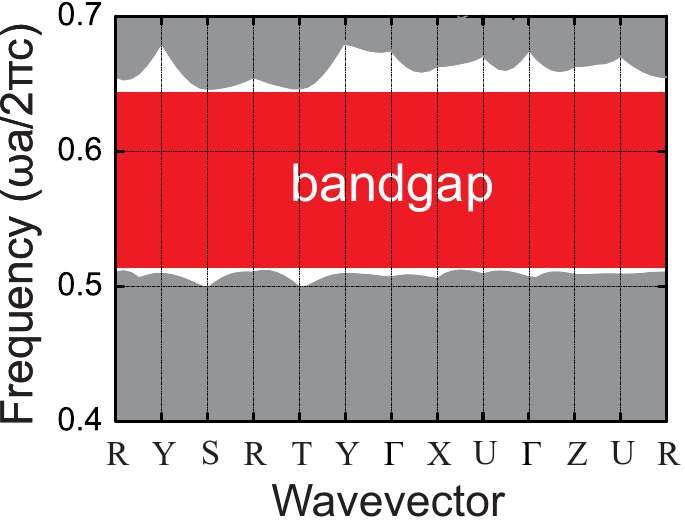}
\caption{\label{BS_3x3x3_no-defect}(Color online) Photonic bandstructure calculated for a perfect inverse woodpile photonic crystal that is defined by a 3$\times$3$\times$3 supercell. Each of the 27 cells was defined with grid resolution of 12$\times$17$\times$12. All pore radii are equal to allow a comparison with a conventional 1$\times$1$\times$1 calculation~\cite{Woldering2009}.
A band gap with a relative width $\frac{{\Delta\omega}}{{\omega}}$~=~23.3\% is found between reduced frequencies $\omega$~=~0.512 and 0.646. For ease of interpretation, all allowed frequencies up to the band gap edges are depicted in grey.}
\end{figure}

In Figure~\ref{BS_3x3x3_no-defect} the calculated bandstructure is shown of a perfect inverse woodpile photonic crystal that was defined by a 3$\times$3$\times$3 supercell. The geometrical properties of the crystal are equal to those used in the conventional 1$\times$1$\times$1 calculations. The band gap edges and band gap width found in this calculation are in excellent agreement with the conventional 1$\times$1$\times$1 calculation, see Table~\ref{Table1}.
Because a supercell is applied, band folding occurs which causes the bands that define the band gap edges to appear flatter.

\section{Calculation of the mode volume \label{appendixVmode}}
\hspace{\parindent}
From the calculated electric-field energy density distributions we have determined mode volumes $V_\mathrm{mode}$ using:

\begin{equation}
V_\mathrm{mode} = \frac{{(\Delta x \Delta y \Delta z)} \sum_\mathrm{ijk} W_\mathrm{ijk}}{W_{\mathrm{max}}},
\label{eq_Vmode}
\end{equation}

with $W_\mathrm{ijk}$ the electric-field energy density in each grid-element with sizes $\Delta x$, $\Delta y$, and $\Delta z$ along each axis, respectively, and $W_{\mathrm{max}}$ the maximum of the electric-field energy density obtained in the entire supercell.
The mode volume is normalized to the cubed wavelength of each resonance $i$: $\frac{V_\mathrm{mode}}{\lambda_\mathrm{i}^3}$.

\clearpage

\begin{thebibliography}{48}
\expandafter\ifx\csname natexlab\endcsname\relax\def\natexlab#1{#1}\fi
\expandafter\ifx\csname bibnamefont\endcsname\relax
  \def\bibnamefont#1{#1}\fi
\expandafter\ifx\csname bibfnamefont\endcsname\relax
  \def\bibfnamefont#1{#1}\fi
\expandafter\ifx\csname citenamefont\endcsname\relax
  \def\citenamefont#1{#1}\fi
\expandafter\ifx\csname url\endcsname\relax
  \def\url#1{\texttt{#1}}\fi
\expandafter\ifx\csname urlprefix\endcsname\relax\def\urlprefix{URL }\fi
\providecommand{\bibinfo}[2]{#2}
\providecommand{\eprint}[2][]{\url{#2}}

\bibitem[{\citenamefont{Vahala}(2003)}]{Vahala2003}
\bibinfo{author}{\bibfnamefont{K.~J.} \bibnamefont{Vahala}},
  \bibinfo{journal}{Nature} \textbf{\bibinfo{volume}{424}},
  \bibinfo{pages}{839} (\bibinfo{year}{2003}).

\bibitem[{\citenamefont{Novotny and Hecht}(2006)}]{Novotny2006}
\bibinfo{author}{\bibfnamefont{L.}~\bibnamefont{Novotny}} \bibnamefont{and}
  \bibinfo{author}{\bibfnamefont{B.}~\bibnamefont{Hecht}},
  \emph{\bibinfo{title}{Principles of Nano-Optics}}
  (\bibinfo{publisher}{Cambridge University Press, Cambridge},
  \bibinfo{year}{2006}).

\bibitem[{\citenamefont{Armani et~al.}(2003)\citenamefont{Armani, Kippenberg,
  Spillane, and Vahala}}]{Armani2003}
\bibinfo{author}{\bibfnamefont{D.~K.} \bibnamefont{Armani}},
  \bibinfo{author}{\bibfnamefont{T.~J.} \bibnamefont{Kippenberg}},
  \bibinfo{author}{\bibfnamefont{S.~M.} \bibnamefont{Spillane}},
  \bibnamefont{and} \bibinfo{author}{\bibfnamefont{K.~J.}
  \bibnamefont{Vahala}}, \bibinfo{journal}{Nature}
  \textbf{\bibinfo{volume}{421}}, \bibinfo{pages}{925} (\bibinfo{year}{2003}).

\bibitem[{\citenamefont{G{\'e}rard}(2003)}]{Gerard2003}
\bibinfo{author}{\bibfnamefont{J.-M.} \bibnamefont{G{\'e}rard}},
  \emph{\bibinfo{title}{\textrm{in} ''Single Quantum Dots" (Topics in Applied
  Physics Series, volume 90)'}} (\bibinfo{publisher}{Springer, Berlin
  Heidelberg}, \bibinfo{year}{2003}).

\bibitem[{\citenamefont{Akahane et~al.}(2003)\citenamefont{Akahane, Asano,
  Song, and Noda}}]{Akahane2003}
\bibinfo{author}{\bibfnamefont{Y.}~\bibnamefont{Akahane}},
  \bibinfo{author}{\bibfnamefont{T.}~\bibnamefont{Asano}},
  \bibinfo{author}{\bibfnamefont{B.~S.} \bibnamefont{Song}}, \bibnamefont{and}
  \bibinfo{author}{\bibfnamefont{S.}~\bibnamefont{Noda}},
  \bibinfo{journal}{Nature} \textbf{\bibinfo{volume}{425}},
  \bibinfo{pages}{944} (\bibinfo{year}{2003}).

\bibitem[{\citenamefont{Hess et~al.}(2003)\citenamefont{Hess, Hermann, and
  Klaedtke}}]{Hess2003pssa}
\bibinfo{author}{\bibfnamefont{O.}~\bibnamefont{Hess}},
  \bibinfo{author}{\bibfnamefont{C.}~\bibnamefont{Hermann}}, \bibnamefont{and}
  \bibinfo{author}{\bibfnamefont{A.}~\bibnamefont{Klaedtke}},
  \bibinfo{journal}{Phys. Stat. Sol. A.} \textbf{\bibinfo{volume}{197}},
  \bibinfo{pages}{605} (\bibinfo{year}{2003}).

\bibitem[{\citenamefont{Farahani et~al.}(2005)\citenamefont{Farahani, Pohl,
  Eisler, and Hecht}}]{Farahani2005}
\bibinfo{author}{\bibfnamefont{J.~N.} \bibnamefont{Farahani}},
  \bibinfo{author}{\bibfnamefont{D.~W.} \bibnamefont{Pohl}},
  \bibinfo{author}{\bibfnamefont{H.~J.} \bibnamefont{Eisler}},
  \bibnamefont{and} \bibinfo{author}{\bibfnamefont{B.}~\bibnamefont{Hecht}},
  \bibinfo{journal}{Phys. Rev. Lett.} \textbf{\bibinfo{volume}{95}},
  \bibinfo{pages}{017402} (\bibinfo{year}{2005}).

\bibitem[{\citenamefont{Taminiau et~al.}(2008)\citenamefont{Taminiau, Stefani,
  Segerink, and {van Hulst}}}]{Taminiau2008}
\bibinfo{author}{\bibfnamefont{T.~H.} \bibnamefont{Taminiau}},
  \bibinfo{author}{\bibfnamefont{F.~D.} \bibnamefont{Stefani}},
  \bibinfo{author}{\bibfnamefont{F.~B.} \bibnamefont{Segerink}},
  \bibnamefont{and} \bibinfo{author}{\bibfnamefont{N.~F.} \bibnamefont{{van
  Hulst}}}, \bibinfo{journal}{Nat. Photonics} \textbf{\bibinfo{volume}{2}},
  \bibinfo{pages}{234} (\bibinfo{year}{2008}).

\bibitem[{\citenamefont{Krioukov et~al.}(2002)\citenamefont{Krioukov, Klunder,
  Driessen, Greve, and Otto}}]{krio2002}
\bibinfo{author}{\bibfnamefont{E.}~\bibnamefont{Krioukov}},
  \bibinfo{author}{\bibfnamefont{D.~J.~W.} \bibnamefont{Klunder}},
  \bibinfo{author}{\bibfnamefont{A.}~\bibnamefont{Driessen}},
  \bibinfo{author}{\bibfnamefont{J.}~\bibnamefont{Greve}}, \bibnamefont{and}
  \bibinfo{author}{\bibfnamefont{C.}~\bibnamefont{Otto}},
  \bibinfo{journal}{Opt. Lett.} \textbf{\bibinfo{volume}{27}},
  \bibinfo{pages}{512} (\bibinfo{year}{2002}).

\bibitem[{\citenamefont{G{\'e}rard et~al.}(1998)\citenamefont{G{\'e}rard,
  Sermage, Gayral, Legrand, Costard, and Thierry-Mieg}}]{Gerard1998}
\bibinfo{author}{\bibfnamefont{J.-M.} \bibnamefont{G{\'e}rard}},
  \bibinfo{author}{\bibfnamefont{B.}~\bibnamefont{Sermage}},
  \bibinfo{author}{\bibfnamefont{B.}~\bibnamefont{Gayral}},
  \bibinfo{author}{\bibfnamefont{B.}~\bibnamefont{Legrand}},
  \bibinfo{author}{\bibfnamefont{E.}~\bibnamefont{Costard}}, \bibnamefont{and}
  \bibinfo{author}{\bibfnamefont{V.}~\bibnamefont{Thierry-Mieg}},
  \bibinfo{journal}{Phys. Rev. Lett.} \textbf{\bibinfo{volume}{81}},
  \bibinfo{pages}{1110} (\bibinfo{year}{1998}).

\bibitem[{\citenamefont{Reithmaier et~al.}(2004)\citenamefont{Reithmaier,
  S\c{e}k, L\"offler, Hoffmann, Kuhn, Reitzenstein, Keldysh, Kulakovskii,
  Reinecke, and Forchel}}]{Reithmaier2004}
\bibinfo{author}{\bibfnamefont{J.~P.} \bibnamefont{Reithmaier}},
  \bibinfo{author}{\bibfnamefont{G.}~\bibnamefont{S\c{e}k}},
  \bibinfo{author}{\bibfnamefont{A.}~\bibnamefont{L\"offler}},
  \bibinfo{author}{\bibfnamefont{C.}~\bibnamefont{Hoffmann}},
  \bibinfo{author}{\bibfnamefont{S.}~\bibnamefont{Kuhn}},
  \bibinfo{author}{\bibfnamefont{S.}~\bibnamefont{Reitzenstein}},
  \bibinfo{author}{\bibfnamefont{L.~V.} \bibnamefont{Keldysh}},
  \bibinfo{author}{\bibfnamefont{V.~D.} \bibnamefont{Kulakovskii}},
  \bibinfo{author}{\bibfnamefont{T.~L.} \bibnamefont{Reinecke}},
  \bibnamefont{and} \bibinfo{author}{\bibfnamefont{A.}~\bibnamefont{Forchel}},
  \bibinfo{journal}{Nature} \textbf{\bibinfo{volume}{432}},
  \bibinfo{pages}{197} (\bibinfo{year}{2004}).

\bibitem[{\citenamefont{Yoshie et~al.}(2004)\citenamefont{Yoshie, Scherer,
  Hendrickson, Khitrova, Gibbs, Rupper, Ell, Shchekin, and
  Deppe}}]{Yoshie2004N}
\bibinfo{author}{\bibfnamefont{T.}~\bibnamefont{Yoshie}},
  \bibinfo{author}{\bibfnamefont{A.}~\bibnamefont{Scherer}},
  \bibinfo{author}{\bibfnamefont{J.}~\bibnamefont{Hendrickson}},
  \bibinfo{author}{\bibfnamefont{G.}~\bibnamefont{Khitrova}},
  \bibinfo{author}{\bibfnamefont{H.~M.} \bibnamefont{Gibbs}},
  \bibinfo{author}{\bibfnamefont{G.}~\bibnamefont{Rupper}},
  \bibinfo{author}{\bibfnamefont{C.}~\bibnamefont{Ell}},
  \bibinfo{author}{\bibfnamefont{O.~B.} \bibnamefont{Shchekin}},
  \bibnamefont{and} \bibinfo{author}{\bibfnamefont{D.~G.} \bibnamefont{Deppe}},
  \bibinfo{journal}{Nature} \textbf{\bibinfo{volume}{432}},
  \bibinfo{pages}{200} (\bibinfo{year}{2004}).

\bibitem[{\citenamefont{Peter et~al.}(2005)\citenamefont{Peter, Senellart,
  Martrou, Lema\^{\i}tre, Hours, G\'{e}rard, and Bloch}}]{Peter2005}
\bibinfo{author}{\bibfnamefont{E.}~\bibnamefont{Peter}},
  \bibinfo{author}{\bibfnamefont{P.}~\bibnamefont{Senellart}},
  \bibinfo{author}{\bibfnamefont{D.}~\bibnamefont{Martrou}},
  \bibinfo{author}{\bibfnamefont{A.}~\bibnamefont{Lema\^{\i}tre}},
  \bibinfo{author}{\bibfnamefont{J.}~\bibnamefont{Hours}},
  \bibinfo{author}{\bibfnamefont{J.~M.} \bibnamefont{G\'{e}rard}},
  \bibnamefont{and} \bibinfo{author}{\bibfnamefont{J.}~\bibnamefont{Bloch}},
  \bibinfo{journal}{Phys. Rev. Lett.} \textbf{\bibinfo{volume}{95}},
  \bibinfo{pages}{067401} (\bibinfo{year}{2005}).

\bibitem[{\citenamefont{Yariv et~al.}(1999)\citenamefont{Yariv, Xu, Lee, and
  Scherer}}]{yari1999}
\bibinfo{author}{\bibfnamefont{A.}~\bibnamefont{Yariv}},
  \bibinfo{author}{\bibfnamefont{Y.}~\bibnamefont{Xu}},
  \bibinfo{author}{\bibfnamefont{R.~K.} \bibnamefont{Lee}}, \bibnamefont{and}
  \bibinfo{author}{\bibfnamefont{A.}~\bibnamefont{Scherer}},
  \bibinfo{journal}{Opt. Lett.} \textbf{\bibinfo{volume}{24}},
  \bibinfo{pages}{711} (\bibinfo{year}{1999}).

\bibitem[{\citenamefont{Yannopapas et~al.}(2002)\citenamefont{Yannopapas,
  Modinos, and Stefanou}}]{Yann2002}
\bibinfo{author}{\bibfnamefont{V.}~\bibnamefont{Yannopapas}},
  \bibinfo{author}{\bibfnamefont{A.}~\bibnamefont{Modinos}}, \bibnamefont{and}
  \bibinfo{author}{\bibfnamefont{N.}~\bibnamefont{Stefanou}},
  \bibinfo{journal}{Phys. Rev. B} \textbf{\bibinfo{volume}{65}},
  \bibinfo{pages}{235201} (\bibinfo{year}{2002}).

\bibitem[{\citenamefont{Yablonovitch}(1987)}]{Yablonovitch1987prl}
\bibinfo{author}{\bibfnamefont{E.}~\bibnamefont{Yablonovitch}},
  \bibinfo{journal}{Phys. Rev. Lett.} \textbf{\bibinfo{volume}{58}},
  \bibinfo{pages}{2059} (\bibinfo{year}{1987}).

\bibitem[{\citenamefont{John}(1987)}]{John1987prl}
\bibinfo{author}{\bibfnamefont{S.}~\bibnamefont{John}}, \bibinfo{journal}{Phys.
  Rev. Lett.} \textbf{\bibinfo{volume}{58}}, \bibinfo{pages}{2486}
  (\bibinfo{year}{1987}).

\bibitem[{\citenamefont{Leistikow et~al.}(2011)\citenamefont{Leistikow, Mosk,
  Yeganegi, Huisman, Lagendijk, and Vos}}]{Leistikow2011PRL}
\bibinfo{author}{\bibfnamefont{M.~D.} \bibnamefont{Leistikow}},
  \bibinfo{author}{\bibfnamefont{A.~P.} \bibnamefont{Mosk}},
  \bibinfo{author}{\bibfnamefont{E.}~\bibnamefont{Yeganegi}},
  \bibinfo{author}{\bibfnamefont{S.~R.} \bibnamefont{Huisman}},
  \bibinfo{author}{\bibfnamefont{A.}~\bibnamefont{Lagendijk}},
  \bibnamefont{and} \bibinfo{author}{\bibfnamefont{W.~L.} \bibnamefont{Vos}},
  \bibinfo{journal}{Phys. Rev. Lett.} \textbf{\bibinfo{volume}{107}},
  \bibinfo{pages}{193903} (\bibinfo{year}{2011}).

\bibitem[{\citenamefont{Yablonovitch et~al.}(1991)\citenamefont{Yablonovitch,
  Gmitter, Meade, Rappe, Brommer, and Joannopoulos}}]{Yablonovitch1991PRL-2}
\bibinfo{author}{\bibfnamefont{E.}~\bibnamefont{Yablonovitch}},
  \bibinfo{author}{\bibfnamefont{T.~J.} \bibnamefont{Gmitter}},
  \bibinfo{author}{\bibfnamefont{R.~D.} \bibnamefont{Meade}},
  \bibinfo{author}{\bibfnamefont{A.}~\bibnamefont{Rappe}},
  \bibinfo{author}{\bibfnamefont{K.~D.} \bibnamefont{Brommer}},
  \bibnamefont{and} \bibinfo{author}{\bibfnamefont{J.~D.}
  \bibnamefont{Joannopoulos}}, \bibinfo{journal}{Phys. Rev. Lett.}
  \textbf{\bibinfo{volume}{67}}, \bibinfo{pages}{3380} (\bibinfo{year}{1991}).

\bibitem[{\citenamefont{{\"O}zbay et~al.}(1995)\citenamefont{{\"O}zbay, Tuttle,
  Sigalas, Soukoulis, and Ho}}]{Ozbay1995}
\bibinfo{author}{\bibfnamefont{E.}~\bibnamefont{{\"O}zbay}},
  \bibinfo{author}{\bibfnamefont{G.}~\bibnamefont{Tuttle}},
  \bibinfo{author}{\bibfnamefont{M.}~\bibnamefont{Sigalas}},
  \bibinfo{author}{\bibfnamefont{C.~M.} \bibnamefont{Soukoulis}},
  \bibnamefont{and} \bibinfo{author}{\bibfnamefont{K.~M.} \bibnamefont{Ho}},
  \bibinfo{journal}{Phys. Rev. B.} \textbf{\bibinfo{volume}{51}},
  \bibinfo{pages}{13961} (\bibinfo{year}{1995}).

\bibitem[{\citenamefont{Villeneuve et~al.}(1996)\citenamefont{Villeneuve, Fan,
  and Joannopoulos}}]{vill1996}
\bibinfo{author}{\bibfnamefont{P.~R.} \bibnamefont{Villeneuve}},
  \bibinfo{author}{\bibfnamefont{S.~H.} \bibnamefont{Fan}}, \bibnamefont{and}
  \bibinfo{author}{\bibfnamefont{J.~D.} \bibnamefont{Joannopoulos}},
  \bibinfo{journal}{Phys. Rev. B.} \textbf{\bibinfo{volume}{54}},
  \bibinfo{pages}{7837} (\bibinfo{year}{1996}).

\bibitem[{\citenamefont{Joannopoulos et~al.}(2008)\citenamefont{Joannopoulos,
  Johnson, Winn, and Meade}}]{Joannopoulos2008}
\bibinfo{author}{\bibfnamefont{J.~D.} \bibnamefont{Joannopoulos}},
  \bibinfo{author}{\bibfnamefont{S.~G.} \bibnamefont{Johnson}},
  \bibinfo{author}{\bibfnamefont{J.~N.} \bibnamefont{Winn}}, \bibnamefont{and}
  \bibinfo{author}{\bibfnamefont{R.~D.} \bibnamefont{Meade}},
  \emph{\bibinfo{title}{Photonic crystals - molding the flow of light}}
  (\bibinfo{publisher}{Princeton University Press, Princeton},
  \bibinfo{year}{2008}), \bibinfo{edition}{2nd} ed.

\bibitem[{\citenamefont{Ashcroft and Mermin}(1976)}]{Ashcroft1976}
\bibinfo{author}{\bibfnamefont{N.~W.} \bibnamefont{Ashcroft}} \bibnamefont{and}
  \bibinfo{author}{\bibfnamefont{N.~D.} \bibnamefont{Mermin}},
  \emph{\bibinfo{title}{Solid State Physics}} (\bibinfo{publisher}{Holt,
  Rinehart, and Winston}, \bibinfo{year}{1976}).

\bibitem[{\citenamefont{Economou}(2010)}]{Economou2010}
\bibinfo{author}{\bibfnamefont{E.~N.} \bibnamefont{Economou}},
  \emph{\bibinfo{title}{The Physics of Solids: Essentials and Beyond}}
  (\bibinfo{publisher}{Springer, New York}, \bibinfo{year}{2010}).

\bibitem[{\citenamefont{Vos and Woldering}(2014)}]{Vos2014Cambridge}
\bibinfo{author}{\bibfnamefont{W.~L.} \bibnamefont{Vos}} \bibnamefont{and}
  \bibinfo{author}{\bibfnamefont{L.~A.} \bibnamefont{Woldering}},
  \emph{\bibinfo{title}{\textrm{in} ''Light Localisation and Lasing: Random and
  Pseudorandom Photonic Structures''}} (\bibinfo{publisher}{Cambridge
  University Press, Cambridge UK}, \bibinfo{year}{2014}).

\bibitem[{\citenamefont{Braun et~al.}(2006)\citenamefont{Braun, Rinne, and
  {Garc\'ia-Santamar\'ia}}}]{brau2006}
\bibinfo{author}{\bibfnamefont{P.~V.} \bibnamefont{Braun}},
  \bibinfo{author}{\bibfnamefont{S.~A.} \bibnamefont{Rinne}}, \bibnamefont{and}
  \bibinfo{author}{\bibfnamefont{F.}~\bibnamefont{{Garc\'ia-Santamar\'ia}}},
  \bibinfo{journal}{Adv. Mater.} \textbf{\bibinfo{volume}{18}},
  \bibinfo{pages}{2665} (\bibinfo{year}{2006}).

\bibitem[{\citenamefont{Yan et~al.}(2007)\citenamefont{Yan, Wang, and
  Zhao}}]{yan2007}
\bibinfo{author}{\bibfnamefont{Q.}~\bibnamefont{Yan}},
  \bibinfo{author}{\bibfnamefont{L.}~\bibnamefont{Wang}}, \bibnamefont{and}
  \bibinfo{author}{\bibfnamefont{X.~S.} \bibnamefont{Zhao}},
  \bibinfo{journal}{Adv. Funct. Mater.} \textbf{\bibinfo{volume}{17}},
  \bibinfo{pages}{3695} (\bibinfo{year}{2007}).

\bibitem[{\citenamefont{Lin et~al.}(1999)\citenamefont{Lin, Fleming, Sigalas,
  Biswas, and Ho}}]{lin1999}
\bibinfo{author}{\bibfnamefont{S.~Y.} \bibnamefont{Lin}},
  \bibinfo{author}{\bibfnamefont{J.~G.} \bibnamefont{Fleming}},
  \bibinfo{author}{\bibfnamefont{M.~M.} \bibnamefont{Sigalas}},
  \bibinfo{author}{\bibfnamefont{R.}~\bibnamefont{Biswas}}, \bibnamefont{and}
  \bibinfo{author}{\bibfnamefont{K.~M.} \bibnamefont{Ho}},
  \bibinfo{journal}{Phys. Rev. B.} \textbf{\bibinfo{volume}{59}},
  \bibinfo{pages}{R15579} (\bibinfo{year}{1999}).

\bibitem[{\citenamefont{Okano et~al.}(2002)\citenamefont{Okano, Chutinan, and
  Noda}}]{Okano2002}
\bibinfo{author}{\bibfnamefont{M.}~\bibnamefont{Okano}},
  \bibinfo{author}{\bibfnamefont{A.}~\bibnamefont{Chutinan}}, \bibnamefont{and}
  \bibinfo{author}{\bibfnamefont{S.}~\bibnamefont{Noda}},
  \bibinfo{journal}{Phys. Rev. B.} \textbf{\bibinfo{volume}{66}},
  \bibinfo{pages}{165211} (\bibinfo{year}{2002}).

\bibitem[{\citenamefont{Ogawa et~al.}(2004)\citenamefont{Ogawa, Imada,
  Yoshimoto, Okano, and Noda}}]{ogaw2004}
\bibinfo{author}{\bibfnamefont{S.~P.} \bibnamefont{Ogawa}},
  \bibinfo{author}{\bibfnamefont{M.}~\bibnamefont{Imada}},
  \bibinfo{author}{\bibfnamefont{S.}~\bibnamefont{Yoshimoto}},
  \bibinfo{author}{\bibfnamefont{M.}~\bibnamefont{Okano}}, \bibnamefont{and}
  \bibinfo{author}{\bibfnamefont{S.}~\bibnamefont{Noda}},
  \bibinfo{journal}{Science} \textbf{\bibinfo{volume}{305}},
  \bibinfo{pages}{227} (\bibinfo{year}{2004}).

\bibitem[{\citenamefont{Tandaechanurat
  et~al.}(2011)\citenamefont{Tandaechanurat, Ishida, Guimard, Nomura, Iwamoto,
  and Arakawa}}]{tand2011}
\bibinfo{author}{\bibfnamefont{A.}~\bibnamefont{Tandaechanurat}},
  \bibinfo{author}{\bibfnamefont{S.}~\bibnamefont{Ishida}},
  \bibinfo{author}{\bibfnamefont{D.}~\bibnamefont{Guimard}},
  \bibinfo{author}{\bibfnamefont{M.}~\bibnamefont{Nomura}},
  \bibinfo{author}{\bibfnamefont{S.}~\bibnamefont{Iwamoto}}, \bibnamefont{and}
  \bibinfo{author}{\bibfnamefont{Y.}~\bibnamefont{Arakawa}},
  \bibinfo{journal}{Nature Photon.} \textbf{\bibinfo{volume}{5}},
  \bibinfo{pages}{91} (\bibinfo{year}{2011}).

\bibitem[{\citenamefont{Tang and Yoshie}(2007)}]{tang2007_2}
\bibinfo{author}{\bibfnamefont{L.}~\bibnamefont{Tang}} \bibnamefont{and}
  \bibinfo{author}{\bibfnamefont{T.}~\bibnamefont{Yoshie}},
  \bibinfo{journal}{Opt. Express} \textbf{\bibinfo{volume}{15}},
  \bibinfo{pages}{17254} (\bibinfo{year}{2007}).

\bibitem[{\citenamefont{Tang and Yoshie}(2011)}]{tang2011}
\bibinfo{author}{\bibfnamefont{L.}~\bibnamefont{Tang}} \bibnamefont{and}
  \bibinfo{author}{\bibfnamefont{T.}~\bibnamefont{Yoshie}},
  \bibinfo{journal}{IEEE J. Quant. Elec.} \textbf{\bibinfo{volume}{47}},
  \bibinfo{pages}{1028} (\bibinfo{year}{2011}).

\bibitem[{\citenamefont{Ferrand et~al.}(2004)\citenamefont{Ferrand, Seekamp,
  Egen, Zentel, Romanov, and Torres}}]{Ferr2004}
\bibinfo{author}{\bibfnamefont{P.}~\bibnamefont{Ferrand}},
  \bibinfo{author}{\bibfnamefont{J.}~\bibnamefont{Seekamp}},
  \bibinfo{author}{\bibfnamefont{M.}~\bibnamefont{Egen}},
  \bibinfo{author}{\bibfnamefont{R.}~\bibnamefont{Zentel}},
  \bibinfo{author}{\bibfnamefont{S.~G.} \bibnamefont{Romanov}},
  \bibnamefont{and} \bibinfo{author}{\bibfnamefont{C.~M.~S.}
  \bibnamefont{Torres}}, \bibinfo{journal}{Microelectron. Eng.}
  \textbf{\bibinfo{volume}{73-74}}, \bibinfo{pages}{362}
  (\bibinfo{year}{2004}).

\bibitem[{\citenamefont{Lee et~al.}(2002)\citenamefont{Lee, Pruzinsky, and
  Braun}}]{Lee2002}
\bibinfo{author}{\bibfnamefont{W.~M.} \bibnamefont{Lee}},
  \bibinfo{author}{\bibfnamefont{S.~A.} \bibnamefont{Pruzinsky}},
  \bibnamefont{and} \bibinfo{author}{\bibfnamefont{P.~V.} \bibnamefont{Braun}},
  \bibinfo{journal}{Adv. Mater.} \textbf{\bibinfo{volume}{14}},
  \bibinfo{pages}{271} (\bibinfo{year}{2002}).

\bibitem[{\citenamefont{Rinne et~al.}(2008)\citenamefont{Rinne,
  {Garc\'ia-Santamar\'ia}, and Braun}}]{rinn2007}
\bibinfo{author}{\bibfnamefont{S.~A.} \bibnamefont{Rinne}},
  \bibinfo{author}{\bibfnamefont{F.}~\bibnamefont{{Garc\'ia-Santamar\'ia}}},
  \bibnamefont{and} \bibinfo{author}{\bibfnamefont{P.~V.} \bibnamefont{Braun}},
  \bibinfo{journal}{Nature Photonics} \textbf{\bibinfo{volume}{2}},
  \bibinfo{pages}{52} (\bibinfo{year}{2008}).

\bibitem[{\citenamefont{Ramanan et~al.}(2008)\citenamefont{Ramanan, Nelson,
  Brzezinski, Braun, and Wiltzius}}]{rama2008}
\bibinfo{author}{\bibfnamefont{V.}~\bibnamefont{Ramanan}},
  \bibinfo{author}{\bibfnamefont{E.}~\bibnamefont{Nelson}},
  \bibinfo{author}{\bibfnamefont{A.}~\bibnamefont{Brzezinski}},
  \bibinfo{author}{\bibfnamefont{P.~V.} \bibnamefont{Braun}}, \bibnamefont{and}
  \bibinfo{author}{\bibfnamefont{P.}~\bibnamefont{Wiltzius}},
  \bibinfo{journal}{Appl. Phys. Lett.} \textbf{\bibinfo{volume}{92}},
  \bibinfo{pages}{173304} (\bibinfo{year}{2008}).

\bibitem[{\citenamefont{Ho et~al.}(1994)\citenamefont{Ho, Chan, Soukoulis,
  Biswas, and {Si}galas}}]{Ho1994SSC}
\bibinfo{author}{\bibfnamefont{K.~M.} \bibnamefont{Ho}},
  \bibinfo{author}{\bibfnamefont{C.~T.} \bibnamefont{Chan}},
  \bibinfo{author}{\bibfnamefont{C.~M.} \bibnamefont{Soukoulis}},
  \bibinfo{author}{\bibfnamefont{R.}~\bibnamefont{Biswas}}, \bibnamefont{and}
  \bibinfo{author}{\bibfnamefont{M.}~\bibnamefont{{Si}galas}},
  \bibinfo{journal}{Solid State Commun.} \textbf{\bibinfo{volume}{89}},
  \bibinfo{pages}{413} (\bibinfo{year}{1994}).

\bibitem[{\citenamefont{Maldovan and Thomas}(2004)}]{Maldovan2004}
\bibinfo{author}{\bibfnamefont{M.}~\bibnamefont{Maldovan}} \bibnamefont{and}
  \bibinfo{author}{\bibfnamefont{E.~L.} \bibnamefont{Thomas}},
  \bibinfo{journal}{Nature Mater.} \textbf{\bibinfo{volume}{3}},
  \bibinfo{pages}{593} (\bibinfo{year}{2004}).

\bibitem[{\citenamefont{Li and Zhang}(2000)}]{Li2000PRB}
\bibinfo{author}{\bibfnamefont{Z.~Y.} \bibnamefont{Li}} \bibnamefont{and}
  \bibinfo{author}{\bibfnamefont{Z.~Q.} \bibnamefont{Zhang}},
  \bibinfo{journal}{Phys. Rev. B.} \textbf{\bibinfo{volume}{62}},
  \bibinfo{pages}{1516} (\bibinfo{year}{2000}).

\bibitem[{\citenamefont{Woldering et~al.}({2009})\citenamefont{Woldering, Mosk,
  Tjerkstra, and Vos}}]{Woldering2009}
\bibinfo{author}{\bibfnamefont{L.~A.} \bibnamefont{Woldering}},
  \bibinfo{author}{\bibfnamefont{A.~P.} \bibnamefont{Mosk}},
  \bibinfo{author}{\bibfnamefont{R.~W.} \bibnamefont{Tjerkstra}},
  \bibnamefont{and} \bibinfo{author}{\bibfnamefont{W.~L.} \bibnamefont{Vos}},
  \bibinfo{journal}{J. Appl. Phys.} \textbf{\bibinfo{volume}{{105}}},
  \bibinfo{pages}{093108} (\bibinfo{year}{{2009}}).

\bibitem[{\citenamefont{Hillebrand et~al.}(2003)\citenamefont{Hillebrand, Senz,
  Hergert, and G{\"o}sele}}]{Hillebrand2003JAP}
\bibinfo{author}{\bibfnamefont{R.}~\bibnamefont{Hillebrand}},
  \bibinfo{author}{\bibfnamefont{S.}~\bibnamefont{Senz}},
  \bibinfo{author}{\bibfnamefont{W.}~\bibnamefont{Hergert}}, \bibnamefont{and}
  \bibinfo{author}{\bibfnamefont{U.}~\bibnamefont{G{\"o}sele}},
  \bibinfo{journal}{J. Appl. Phys.} \textbf{\bibinfo{volume}{94}},
  \bibinfo{pages}{2758} (\bibinfo{year}{2003}).

\bibitem[{\citenamefont{Schilling et~al.}(2005)\citenamefont{Schilling, White,
  Scherer, Stupian, Hillebrand, and G{\"o}sele}}]{Schilling2005APL}
\bibinfo{author}{\bibfnamefont{J.}~\bibnamefont{Schilling}},
  \bibinfo{author}{\bibfnamefont{J.}~\bibnamefont{White}},
  \bibinfo{author}{\bibfnamefont{A.}~\bibnamefont{Scherer}},
  \bibinfo{author}{\bibfnamefont{G.}~\bibnamefont{Stupian}},
  \bibinfo{author}{\bibfnamefont{R.}~\bibnamefont{Hillebrand}},
  \bibnamefont{and}
  \bibinfo{author}{\bibfnamefont{U.}~\bibnamefont{G{\"o}sele}},
  \bibinfo{journal}{Appl. Phys. Lett.} \textbf{\bibinfo{volume}{86}},
  \bibinfo{pages}{011101} (\bibinfo{year}{2005}).

\bibitem[{\citenamefont{Tjerkstra et~al.}(2011)\citenamefont{Tjerkstra,
  Woldering, {van den Broek}, Roozeboom, Setija, and Vos}}]{Tjerkstra2011JVST}
\bibinfo{author}{\bibfnamefont{R.~W.} \bibnamefont{Tjerkstra}},
  \bibinfo{author}{\bibfnamefont{L.~A.} \bibnamefont{Woldering}},
  \bibinfo{author}{\bibfnamefont{J.~M.} \bibnamefont{{van den Broek}}},
  \bibinfo{author}{\bibfnamefont{F.}~\bibnamefont{Roozeboom}},
  \bibinfo{author}{\bibfnamefont{I.~D.} \bibnamefont{Setija}},
  \bibnamefont{and} \bibinfo{author}{\bibfnamefont{W.~L.} \bibnamefont{Vos}},
  \bibinfo{journal}{J. Vac. Sci. Technol. B} \textbf{\bibinfo{volume}{29}},
  \bibinfo{pages}{061604} (\bibinfo{year}{2011}).

\bibitem[{\citenamefont{{van den Broek} et~al.}(2012)\citenamefont{{van den
  Broek}, Woldering, Tjerkstra, Segerink, Setija, and Vos}}]{Broek2012AFM}
\bibinfo{author}{\bibfnamefont{J.~M.} \bibnamefont{{van den Broek}}},
  \bibinfo{author}{\bibfnamefont{L.~A.} \bibnamefont{Woldering}},
  \bibinfo{author}{\bibfnamefont{R.~W.} \bibnamefont{Tjerkstra}},
  \bibinfo{author}{\bibfnamefont{F.~B.} \bibnamefont{Segerink}},
  \bibinfo{author}{\bibfnamefont{I.~D.} \bibnamefont{Setija}},
  \bibnamefont{and} \bibinfo{author}{\bibfnamefont{W.~L.} \bibnamefont{Vos}},
  \bibinfo{journal}{Adv. Funct. Mater.} \textbf{\bibinfo{volume}{22}},
  \bibinfo{pages}{25} (\bibinfo{year}{2012}).

\bibitem[{\citenamefont{Huisman et~al.}(2011)\citenamefont{Huisman, Nair,
  Woldering, Leistikow, Mosk, and Vos}}]{Huisman2011PRB}
\bibinfo{author}{\bibfnamefont{S.~R.} \bibnamefont{Huisman}},
  \bibinfo{author}{\bibfnamefont{R.~V.} \bibnamefont{Nair}},
  \bibinfo{author}{\bibfnamefont{L.~A.} \bibnamefont{Woldering}},
  \bibinfo{author}{\bibfnamefont{M.~D.} \bibnamefont{Leistikow}},
  \bibinfo{author}{\bibfnamefont{A.~P.} \bibnamefont{Mosk}}, \bibnamefont{and}
  \bibinfo{author}{\bibfnamefont{W.~L.} \bibnamefont{Vos}},
  \bibinfo{journal}{Phys. Rev. B} \textbf{\bibinfo{volume}{83}},
  \bibinfo{pages}{205313} (\bibinfo{year}{2011}).

\bibitem[{\citenamefont{Johnson and Joannopoulos}(2001)}]{Johnson2001aa}
\bibinfo{author}{\bibfnamefont{S.~G.} \bibnamefont{Johnson}} \bibnamefont{and}
  \bibinfo{author}{\bibfnamefont{J.~D.} \bibnamefont{Joannopoulos}},
  \bibinfo{journal}{Opt. Express} \textbf{\bibinfo{volume}{8}},
  \bibinfo{pages}{173} (\bibinfo{year}{2001}).

\bibitem[{\citenamefont{Meade et~al.}(1993)\citenamefont{Meade, Rappe, Brommer,
  Joannopoulos, and Alerhand}}]{Meade1993}
\bibinfo{author}{\bibfnamefont{R.~D.} \bibnamefont{Meade}},
  \bibinfo{author}{\bibfnamefont{A.~M.} \bibnamefont{Rappe}},
  \bibinfo{author}{\bibfnamefont{K.~D.} \bibnamefont{Brommer}},
  \bibinfo{author}{\bibfnamefont{J.~D.} \bibnamefont{Joannopoulos}},
  \bibnamefont{and} \bibinfo{author}{\bibfnamefont{O.~L.}
  \bibnamefont{Alerhand}}, \bibinfo{journal}{Phys. Rev. B.}
  \textbf{\bibinfo{volume}{48}}, \bibinfo{pages}{8434} (\bibinfo{year}{1993}).

\end{thebibliography}

\end{document}